\documentclass[]{aa}
\pdfoutput=1                            
\usepackage[varg]{txfonts}

\usepackage{amsmath}
\usepackage{amsfonts}
\usepackage{amssymb}
\usepackage{wasysym}
\usepackage{graphicx}

\usepackage{natbib}
\usepackage{epsfig}
\usepackage{xspace}
\usepackage{booktabs}
\usepackage{dcolumn}
\usepackage[para]{footmisc}
\usepackage{placeins}
\usepackage{multicol}
\usepackage{color}

\usepackage[switch]{lineno}

\usepackage[utf8]{inputenc}
\usepackage[english]{babel}

\newcommand{\ie}{i.e.}

\newcommand{\emth}[1]{\ensuremath{#1}\xspace}

\newcommand{\covmat}{\emth{\boldsymbol{\Sigma}}}

\newcommand{\flux}{\ensuremath{\mathbf{f}}\xspace}

\newcommand{\npb}{\emth{n_\mathrm{pb}}}

\newcommand{\gcm}{\emth{\mathrm{g\,cm^{-3}}}}
\newcommand{\mjup}{\emth{\mathrm{M_{Jup}}}} 	
\newcommand{\rjup}{\emth{\mathrm{R_{Jup}}}} 	

\newcommand{\logg}{log {\it g}}

\newcommand{\llh}{\ensuremath{\ln P}}

\newcommand{\pvec}{\ensuremath{\boldsymbol{\theta}}\xspace}
\newcommand{\Dlc}{\ensuremath{\vec{D_{\mathrm{LC}}}}\xspace}

\newcommand{\Dld}{\ensuremath{\vec{D_{\mathrm{LD}}}}\xspace}

\newcommand{\UP}[1]{\ensuremath{\mathcal{U}(#1)}\xspace}
\newcommand{\NP}[1]{\ensuremath{\mathcal{N}(#1)}\xspace}

\newcommand{\rstar}{\emth{R_\star}}

\newcommand{\pbg}{\emth{g'}}
\newcommand{\pbr}{\emth{r'}}
\newcommand{\pbi}{\emth{i'}}
\newcommand{\pbz}{\emth{z'}}
\newcommand{\pbj}{\emth{J}}
\newcommand{\pbh}{\emth{H}}
\newcommand{\pbk}{\emth{K}}

\newcommand{\dsw}{\textsc{W}\xspace}
\newcommand{\dsnb}{\textsc{NB}\xspace}
\newcommand{\dsk}{\textsc{K~I}\xspace}
\newcommand{\dsna}{\textsc{Na~I}\xspace}

\newcommand{\mdr}{\textsc{DIR}\xspace}
\newcommand{\mww}{\textsc{DWW}\xspace}
\newcommand{\mwr}{\textsc{DWR}\xspace}

\newcommand{\tmodel}{\ensuremath{\mathcal{T}}\xspace}

\newcommand{\pytransit}{\textsc{PyTransit}\xspace}
\newcommand{\ldtk}{\textsc{LDTk}\xspace}
\newcommand{\george}{\textsc{George}}

\newcommand{\changed}[1]{\textcolor{red}{#1}}
\renewcommand{\changed}[1]{{#1}}

\begin{document}

\title{The GTC exoplanet transit spectroscopy survey}
\subtitle{VIII. Flat transmission spectrum for the warm gas giant WASP-80b}

\author{
 Parviainen, H.\inst{\ref{iiac},\ref{iull},\ref{ioxford}}
 \and Pall\'e, E.\inst{\ref{iiac},\ref{iull}}
 \and Chen, G.\inst{\ref{iiac},\ref{iull},\ref{ipmo}}
 \and Nortmann, L.\inst{\ref{iiac},\ref{iull}}
 \and Murgas, F.\inst{\ref{iiac},\ref{iull}}
  \and Nowak, G.\inst{\ref{iiac},\ref{iull}} 
 \and Aigrain, S.\inst{\ref{ioxford}}
 \and Booth, A.\inst{\ref{ileeds}}
 \and Abazorius, M.\inst{\ref{ioxford}}
 \and Iro, N.\inst{\ref{iham}} 
}

\institute{
Instituto de Astrof\'isica de Canarias (IAC), E-38200 La Laguna, Tenerife, Spain\label{iiac}
\and Dept. Astrof\'isica, Universidad de La Laguna (ULL), E-38206 La Laguna, Tenerife, Spain\label{iull}
\and  Sub-department of Astrophysics, Department of Physics, University of Oxford, Oxford, OX1 3RH, UK\label{ioxford}
\and Key Laboratory of Planetary Sciences, Purple Mountain Observatory, Chinese Academy of Sciences, Nanjing 210008, China\label{ipmo}
\and School of Physics and Astronomy, University of Leeds, Leeds LS2 9JT, UK\label{ileeds}
\and Theoretical Meteorology group, Klimacampus, University of Hamburg, Grindelberg 5, 20144 Hamburg, Germany\label{iham}
}
\date{Received ; accepted}

\abstract{}{We set out to study the atmosphere of WASP-80b, a warm inflated gas giant with an equilibrium temperature
		of $\sim$800~K, using ground-based transmission spectroscopy covering the spectral range from 520~to~910~nm.
		The observations allow us to probe the existence and abundance of K and Na in WASP-80b's atmosphere, existence of
		high-altitude clouds, and Rayleigh-scattering in the blue end of the spectrum.}
	{We observed two spectroscopic time series of WASP-80b transits with the OSIRIS spectrograph 
		installed in the Gran Telescopio CANARIAS, and use the observations to estimate the planet's
		transmission spectrum between 520~nm and 910~nm in 20~nm-wide passbands, and around the K~I and Na~I resonance 
		doublets in 6~nm-wide passbands. 
		We model three previously published broadband datasets consisting of 27 light curves  jointly prior to the transmission
		spectroscopy analysis in order to obtain improved prior estimates for the planet's orbital parameters, average radius ratio, and stellar 
		density. The parameter posteriors from the broadband analysis are used to set informative priors on the 
		transmission spectroscopy analysis. The final transmission spectroscopy analyses are carried out jointly for the two nights
		using a divide-by-white approach to remove the common-mode systematics and Gaussian processes to model the residual 
		wavelength-dependent systematics.}
  	{We recover a flat transmission spectrum with no evidence of Rayleigh scattering or K~I or Na~I absorption, and obtain an improved
  		system characterisation as a by-product of the broadband- and GTC-dataset modelling. The transmission spectra estimated separately 
  		from the two observing runs are consistent with each other, as are the transmission spectra estimated using either a parametric
  		or nonparametric systematics models. The flat transmission spectrum favours an atmosphere model with high-altitude clouds 
  		over cloud-free models with stellar or sub-stellar metallicities.} 
  	{Our results disagree with the recently published discovery of strong K~I absorption in WASP-80b's atmosphere based on ground-based
  		transmission spectroscopy with FORS2 at VLT. }


\titlerunning{}
\authorrunning{}

\maketitle

\section{Introduction}
\label{sec:introduction}

Transmission spectroscopy allows us to probe the existence and abundance of atmospheric species in 
the atmospheres of transiting extrasolar planets \citep{Seager2000,Brown2001a}. The method requires 
a high observing precision, which has made the space-based studies most successful in finding 
significant features in the transmission spectra \citep{Charbonneau2002,Sing2011,Gibson2012}, but the 
developments in instrumentation, observing techniques, and data analysis methods have also enabled
 transmission spectroscopy studies to be carried out successfully using ground-based telescopes.

The signal of interest -- variations in the effective planetary radius as a function of wavelength -- is 
minute, corresponding to changes of $\sim0.01\%$ in the observed transit depth and $\sim0.1\%$ in 
the effective planet-star radius ratio. Further complications arise from possible high-altitude clouds, 
which can mask any atmospheric extinction features, leading to a flat transmission spectrum 
\citep{Kreidberg2013,Berta2011}, and from the fact that atmospheric extinction is not 
the only source of  wavelength-dependent features in transmission spectra. Both instrumental and
astrophysical sources, such as host star's spots \citep{Ballerini2012} and plages \citep{Oshagh2014}, 
flux contamination from a possible unresolved source, and incorrectly accounted for stellar limb darkening, 
can all imprint features that can be difficult to disentangle from the atmospheric signal.

Notwithstanding the complications, ground-based transmission spectroscopy has been used 
successfully to identify features attributed to absorption in planetary atmospheres.
Simultaneous measurements of the target star and several comparison stars -- a process similar to 
relative photometry~\citep{Bean2010a,Gibson2012a} -- the use of Gaussian processes have 
facilitated the robust modelling of systematics~\citep{Gibson2011a,Roberts2013,Rasmussen2006},
and the use of Bayesian inference methods has allowed for realistic uncertainty estimation that is
crucial when assessing the true significance of the identified transmission spectrum features.

We report a ground-based transmission spectroscopy study of WASP-80b \citep{Triaud2013}. We have observed 
spectroscopic time series of two WASP-80b transits with the OSIRIS spectrograph \citep[Optical 
System for Imaging and low-Intermediate-Resolution Integrated Spectroscopy;][]{Sanchez2012} 
installed in the 10.4~m Gran Telescopio CANARIAS (GTC) on La Palma, Spain. The observations cover 
the spectral range from 520~to~910~nm, probing the planet atmosphere for a possible Rayleigh 
scattering signal in the blue end of the spectrum, and the visible-light extinction features of the 
K~I and Na~I resonance doublets at 767~nm and 589.4~nm, respectively.

\object{WASP-80b} \citep[][see also Table~\ref{tbl:star}]{Triaud2013,Mancini2014,Fukui2014a,Triaud2015}, a 
warm gas giant orbiting a bright (V=11.87) late-K / early-M dwarf on a 3.07~d orbit, was identified 
as a promising target for transmission spectroscopy from its discovery. The planet has a low 
surface gravity ($M_\mathrm{p} = 0.56\;\mjup$, $R_\mathrm{p} = 0.99 \rjup$, $g = 14.34$~ms$^{-2}$, 
\citealt{Mancini2014}), and its large radius ratio leads to $\sim$3\% deep transits, which, combined 
with the brightness of its host star, enhance our abilities to detect any possible transmission 
spectrum features. WASP-80b is a warm gas giant with an equilibrium temperature $\sim$800~K 
\citep{Triaud2013,Mancini2014}. This very likely places the planet into the pL class (no temperature 
inversion) in the classification by \citet{Fortney2008}. The main spectroscopic features in the 
visible passband for pL class planets are expected to be from Rayleigh scattering and K~I 
and Na~I resonance doublet absorption, of which  K~I absorption detection was recently claimed by 
\citet{Sedaghati2017} based on transmission spectroscopy analysis carried out with the FORS2 spectrograph 
installed in the VLT.

\begin{table}[t]    
  \caption{Identifiers for WASP-80 with its coordinates and magnitudes \changed{(SIMBAD, retrieved 2017-07-04)}.}
  \centering
  \begin{tabular*}{\columnwidth}{@{\extracolsep{\fill}} llr}
  \toprule\toprule
  \multicolumn{1}{l}{\emph{Main identifiers}}     \\
  \midrule              
  GSC~ID          & 05165-00481         \\
  2MASS~ID        & J20124017-0208391   \\
  WASP~ID         & J201240.26-020838.2 \\
  \midrule               
  \multicolumn{2}{l}{\emph{Equatorial coordinates}}     \\
  \midrule            
  RA \,(J2000)      & $20^h\,12^m\,40\fs1656$            \\
  Dec (J2000)       & $-2\degr\,08\arcmin\,39\farcs194$  \\
  \midrule              
  \multicolumn{3}{l}{\emph{Magnitudes}} \\
  \midrule              
  \centering
  Filter & Magnitude       & Error  \\
  \midrule                
  $B$  & 12.810 & - \\
  $V$  & 11.939 & - \\
  $R$  & 11.510 & - \\
  $I$  & 10.279 & 0.105 \\
  $J$  &  9.218 & 0.023 \\
  $H$  &  8.513 & 0.026 \\
  $K$  &  8.351 & 0.022 \\
  \bottomrule
  \end{tabular*}
  \label{tbl:star}  
\end{table}

Our study consists of two main analyses:
\begin{enumerate}
	\item joint analysis of 27 previously published broadband transit light curves observed in 7 passbands (broadband dataset),
	\item joint analysis of the two GTC-observed spectroscopic transit time series (transmission spectroscopy dataset),
\end{enumerate}
which both consist of a set of analyses with different prior assumptions and modelling approaches
carried out to ensure the robustness of the final results.

The broadband dataset analysis is carried out to obtain improved estimates for the planet's broadband radius ratios, 
orbital parameters, and stellar density (system parameters). The marginal parameter posteriors from the broadband dataset analysis 
are then used as priors in the GTC-data analysis. The GTC transmission spectroscopy starts with a direct-modelling
analysis with a flexible Gaussian-process-based systematics model to further constrain the system parameters,
and the final transmission spectroscopy uses a divide-by-white approach with either a parametric or nonparametric
residual systematics model.

This paper is divided roughly into three sections. We outline the numerical methods and the 
generic equations for the calculation of posterior probability densities in  §\ref{sec:theory}.
We continue in §\ref{sec:ext} by carrying out a detailed joint modelling of three priorly observed 
broadband datasets described in \citet{Triaud2013}, \citet{Mancini2014}, and \citet{Fukui2014a}. The 
datasets cover 27 transit light curves observed in \pbg, \pbr, \pbi, $I$, \pbz, \pbj, \pbh, and \pbk. 
We describe the GTC observations and data reduction in §\ref{sec:data} , detail the analysis in §\ref{sec:gtc_analysis},
present the transmission spectroscopy results in §\ref{sec:results}, and discuss the results in  §\ref{sec:discussion}.
Finally, we conclude the paper in §\ref{sec:conclusions}.

The raw data are publicly available from Zenodo and GTC data archive, and the whole analysis 
with reduced data is available from  GitHub
\begin{quote}
	\url{github.com/hpparvi/Parviainen-2017-WASP-80b}
\end{quote}
as an easy-to-follow set of IPython notebooks and Python codes to help with the reproducibility 
of the study.

\section{Numerical methods and theory} 
\label{sec:theory}
\subsection{Overview}
\label{sec:theory.overview}

We use a fully Bayesian approach to transmission spectroscopy: our parameter estimates are based on the marginal
posterior densities derived from a joint posterior density estimated with Markov Chain Monte Carlo (MCMC) sampling,
and the marginal parameter posteriors from the broadband dataset analysis are used as priors in the GTC transmission 
spectroscopy analysis. 

Our datasets consist of light curves observed either photometrically, or constructed from  
spectroscopic observations.  A light curve is modelled as a product of a baseline and transit model, where the
combined model is parametrised with a parameter vector \pvec. When modelling multiple light curves jointly,
the parameter vector is divided into parameters shared between all the light curves, passband-specific parameters
shared between light curves observed in the same passband, observing-run-specific parameters, and light-curve-specific parameters. Especially, 
the parameters defining the planetary orbit (zero epoch, orbital period, impact parameter, and stellar density)
are shared between all the light curves included into the analysis, and thus the likelihoods from all the light curves
contribute to the parameter posteriors. The planet-star radius ratios and stellar limb darkening coefficients are considered
passband-dependent, but observing-run-independent parameters (although this is not strictly true, since spots and plages, 
whether occulted or not, have an effect on the transit depths). That is, the light curves observed in a given passband all contribute 
to the radius ratio and limb darkening posteriors for the passband. Finally, the parameters defining the baseline and noise properties 
are considered either observing-run- or light-curve-specific, \ie, they depend both on the passband and observing run (this depends on the
specific modelling approach, as detailed later).

Joint modelling leads to relatively high-dimensional models. The number of free parameters in the analyses presented here
varies from 10 to~$\sim$250. However, the approach allows us to utilise the data fully. Simultaneous modelling of different observing 
runs reduces our sensitivity on systematics, and simultaneous multiband analysis  reduces the degeneracies between the
estimated radius ratios, orbital impact parameter, and stellar limb darkening.

The parameter posteriors are estimated as a two-step process. First, a population-based global optimisation method 
(Differential evolution implemented in \textsc{PyDE}) is used to obtain a parameter vector population that is clumped 
close to the global posterior maximum. The parameter vector population is then used to initialise the \textsc{emcee} 
Markov Chain Monte Carlo (MCMC) sampler \citep{Foreman-Mackey2012,Goodman2010}, which is used to create a 
sample of parameter vectors drawn from the model posterior (see the analysis-specific sections for practical details).

The transit model uses the quadratic limb darkening formalism by \citet{Mandel2002}, and is 
calculated using \pytransit \citep{Parviainen2015}. \pytransit contains optimisations to compute a transit in multiple passbands 
with only a minor additional computational cost to the computation of a single passband transit, which reduces the computational
burden due to the joint modelling approach.

The analyses have been carried out both with and without \changed{\ldtk-based} constraints on the stellar limb darkening. 
Quadratic limb darkening is parametrised using the parametrisation presented in \citet{Kipping2013b}, which
is aimed for efficient sampling of the physically allowed limb darkening coefficient space.
When the limb darkening is not constrained, we marginalise over the whole limb darkening parameter space 
allowed by the data. When the limb darkening is constrained, it is done by fitting the observational 
data jointly with the limb darkening profiles created using the \ldtk-package \citep{Parviainen2016},
\changed{and by marginalising over the limb darkening coefficients allowed by the stellar density profiles}.
\ldtk uses \textsc{PHOENIX}-calculated stellar atmosphere library by \citet{Husser2013} to construct 
limb darkening profiles with the uncertainties in the stellar properties propagated into the 
uncertainties in the limb darkening profiles.

We also repeat the analyses for different systematics-modelling approaches, for separate subsets of
data, and with synthethic mock data, and using the target star alone without dividing by the comparison
star, to test the reliability of our approach. 

Unless otherwise specified, the parameter point estimates correspond to posterior medians, and 
 the uncertainties correspond to the central 68\%  posterior intervals. We do not plot point parameter
 estimates (these are listed in tables), but prefer to show either the posterior distributions or limits based
 on central posterior intervals.
 
The analyses rely on Python- and Fortran-based code utilising \textsc{SciPy}, \textsc{NumPy} 
\citep{VanderWalt2011}, 
\textsc{IPython} \citep{Perez2007}, \textsc{Pandas} \citep{Mckinney2010}, \textsc{matplotlib} \citep{Hunter2007}, 
\textsc{seaborn},$\!$\footnote{\url{http://stanford.edu/~mwaskom/software/seaborn}} 
\textsc{PyFITS},$\!$\footnote{PyFITS is a product of the Space Telescope Science Institute, which is operated 
by AURA for NASA} and \textsc{F2PY} \citep{Peterson2009}. The transits were modelled with 
\pytransit\footnote{Freely available from \url{https://github.com/hpparvi/PyTransit}} 
\citep{Parviainen2015},  the limb darkening computations were carried out with 
\ldtk$\!$\footnote{Available from \url{https://github.com/hpparvi/ldtk}} \citep{Parviainen2016}, global 
optimisation was carried out with \textsc{PyDE},$\!$\footnote{Available from 
\url{https://github.com/hpparvi/PyDE}} the MCMC sampling was carried out with emcee 
\citep{Foreman-Mackey2012,Goodman2010}, and the Gaussian processes were computed using 
\george\footnote{Available from \url{https://dan.iel.fm/george}} \citep{Ambikasaran2014}.

\subsection{Posteriors and likelihoods}
\label{sec:analysis:posteriors}

The unnormalised log posterior density for a dataset consisting of $n_\mathrm{lc}$ light curves 
observed in $n_\mathrm{pb}$ passbands is
\begin{equation}
\ln P(\pvec|D) = \ln P(\pvec) + \sum_i^{n_\mathrm{lc}} \llh(\Dlc_{,i}|\pvec) + 
\sum_i^{n_\mathrm{pb}} \llh(\Dld_{,i}|\pvec),
\end{equation}
where \pvec is the parameter vector encapsulating all the model parameters, $\ln P(\pvec)$ is the 
log prior, $\Dlc$ are the light curves, $\ln P(\Dlc|\pvec)$ is 
the log likelihood for the photometry, $\Dld$ are the theoretical limb darkening profiles
calculated by \ldtk, and $\ln P(\Dld|\pvec)$ is the log likelihood for the limb darkening profile.

Assuming that the uncertainties (noise) in the observations are normally distributed, we can write 
the log likelihood for the data $\vec{D}$ in vector form as
\begin{equation}
 \llh(\vec{D}|\pvec) = -\frac{1}{2} \left( n_\mathrm{D} \ln 2\pi +\ln|\covmat| +\vec{r}^\mathrm{T} 
\covmat^{-1} 
\vec{r}\right),
 \label{eq:lnlikelihood_gn}
\end{equation}
where $n_\mathrm{D}$ is the number of datapoints, $\vec{r}$ is the residual vector, and $\covmat$ is 
the covariance matrix. 
If the noise is white (uncorrelated), the covariance matrix is diagonal, and the likelihood can be 
written out explicitly in scalar form as
\begin{equation}
  \llh(\vec{D}|\pvec) = -\frac{1}{2}\left(n_\mathrm{D}\ln2\pi +\sum_j^{n_\mathrm{D}} \ln 
\sigma_{\mathrm{j}}^2 + \sum_{j=1}^{n_\mathrm{D}} 
\frac{\vec{r}^2}{2\sigma_{\mathrm{j}}^2} \right ),
\end{equation}
where $\sigma_{\mathrm{j}}$ is the uncertainty of the $j$th datapoint. If the per-point uncertainty 
does not vary significantly, this equation can be simplified further into
\begin{equation}
  \llh(\vec{D}|\pvec) = -\frac{n_\mathrm{D}}{2} \ln 2\pi\sigma_{\mathrm{i}}^2 - 
\frac{1}{2} \sum_{j=1}^{n_\mathrm{D}} 
\frac{\vec{r}^2}{2\sigma_{\mathrm{j}}^2}.
\label{eq:lnlikelihood_wn}
\end{equation}

If the noise (here used to describe the leftover variation not explained by the parametric transit 
and baseline models) is not white, the covariance matrix will have off-diagonal elements, and the 
matrix needs to be inverted for the likelihood evaluation. This is the case when the noise is
presented as a Gaussian process \citep{Rasmussen2006,Gibson2011a,Roberts2013}. The 
covariance matrix \covmat in Eq.~\eqref{eq:lnlikelihood_gn} is now
\begin{equation}
 \covmat =  \vec{K}(\vec{x},\vec{x}) + \sigma^2\vec{I},
\end{equation}
where $\vec{K}(\vec{x},\vec{x})$ is defined by a covariance function (also known as a covariance
kernel).   

We describe the likelihood and covariance functions separately for the broadband dataset analysis and 
GTC transmission spectroscopy in Sects.~\ref{sec:ext} and \ref{sec:gtc_analysis}, respectively, 
since the two analyses use slightly different modelling approaches.

\section{Broadband dataset analysis}
\label{sec:ext}
\subsection{Overview}
\label{sec:ext.overview}

We estimate the posterior densities for the WASP-80b orbital parameters based on the three 
datasets observed by \citet{Mancini2014}, \citet{Triaud2013}, and \citet{Fukui2014a}, abbreviated
from herein as M14, T13, and F14, respectively. 

The datasets contain 27 light curves (Fig.~\ref{fig:ext_modelled}) observed in \pbg, \pbr, \pbi, 
$I$, \pbz, \pbj, \pbh, and \pbk. We model all the light curves jointly, and, in contrast to 
\citet{Mancini2014}, include the M14 GROND J, H, and K light curves. We also simplify the analysis 
slightly by merging the $I$ and $i'$ passbands. 

The analysis is carried out assuming either a parametric or nonparametric systematics model (white or red noise), 
constant or wavelength-dependent radius ratio, and with and without \ldtk to constrain the stellar limb darkening, 
which leads to eight separate analysis sets defined in Table~\ref{tbl:ext_analyses}. The T13 and 
M14 datasets, as obtained from VizieR, were detrended by their original authors, and do not 
include other information than the observation time, flux, and error estimates. The F14 dataset was 
kindly provided by the author without detrending, and included all the auxiliary information 
(such as the airmass, and x- and y-centroid shifts for each exposure) used in the original analysis 
presented in \citet{Fukui2014a}.

\begin{table}[t]    
  \caption{Broadband analysis runs for the external datasets. The white-noise runs include only the
\citet[][T]{Triaud2013} and \citet[][M]{Mancini2014} datasets, while the red-noise runs also include the light curves
by \citet[][F]{Fukui2014a}. The constant and varying radius ratios mark whether the radius ratio 
was allowed to vary from passband to passband, or whether it was assumed to be wavelength-independent.}
  \centering
  \begin{tabular*}{\columnwidth}{@{\extracolsep{\fill}} lllll}
  \toprule\toprule
Run name & Noise & Radius ratio & LDTk & Datasets\\
\midrule
ckwn       & White & Constant & No  & TM \\
ckwn\_ldtk & White & Constant & Yes & TM \\
vkwn       & White & Varying  & No  & TM \\
vkwn\_ldtk & White & Varying  & Yes & TM \\
ckrn       & Red   & Constant & No  & TMF\\
ckrn\_ldtk & Red   & Constant & Yes & TMF\\
vkrn       & Red   & Varying  & No  & TMF\\
vkrn\_ldtk & Red   & Varying  & Yes & TMF\\
\bottomrule
\end{tabular*}
\label{tbl:ext_analyses}
\end{table}

\begin{table}[t]    
  \caption{Marginal posterior medians from the external data broadband analysis (run \texttt{ckrn\textunderscore{}ldtk}
  with systematics modelled using GPs, wavelength-independent radius ratio, and limb darkening
  constrained with \textsc{LDTk}). The uncertainties correspond to the central 68\% posterior intervals.}
  \centering
  \begin{tabular*}{\columnwidth}{@{\extracolsep{\fill}} llll}
  \toprule\toprule
Parameter & Units & Posterior median & Uncertainty \\
\midrule
Zero epoch       & BJD& 2456125.41759  & 9.3e-05 \\
Period           &days&       3.067860 & 8.6e-07 \\
Impact parameter &    &       0.215    & 2.1e-02 \\
Stellar density  &\gcm&       4.090   & 5.5e-02 \\
Radius ratio     &    &       0.1715   & 3.2e-04 \\
\bottomrule
\end{tabular*}
\label{tbl:ext_results}
\end{table}
 
\begin{table}[t]
  \caption{Broadband radius ratio estimates and their uncertainties from the  \texttt{vkrn\textunderscore{}ldtk} run.}
  \begin{tabular*}{\columnwidth}{@{\extracolsep{\fill}} cccc}
  \toprule\toprule
  g' & r' & i' & z' \\
  \midrule
  0.16936 & 0.17115 & 0.17185 & 0.17131 \\ 
  (0.00071) & (0.00045) & (0.00038) & (0.00060) \\  
  \\
  \midrule
  J & H & K & \\
  \midrule
  0.16985 & 0.17238 &  0.17195 & \\
  (0.00097) & (0.00103) & (0.00120) & \\
  \bottomrule
\end{tabular*}
\label{tbl:ext_ks}
\end{table}
 
We used the T13 and M14 datasets for an initial modelling with a parametric systematics model and white
additive noise, and include the F14 dataset in the final runs using nonparametric systematics model. The nonparametric
systematics model represents the systematics (and white noise)  as a Gaussian process (GP)
with time as the only covariate for the T13 and M14 datasets, and with time, airmass, x-shift, 
and y-shift as covariates for the F14 dataset. We do not marginalise over the GP 
hyperparameters in the broadband data analysis, but fix them to the values fitted from the white-noise run residuals.

As described in Sect.~\ref{sec:theory.overview}, the parameter estimation starts with a parameter vector population that fills  the 
prior space uniformly. A differential evolution (DE) optimisation is used to clump the population 
close to the global posterior maximum (the number of DE iterations depending slightly on the run, 
but is usually close to 1000), after which MCMC sampling is carried out using \textsc{emcee}. The 
sampler is run for 10$\,$000 iterations, which yields 12000 independent posterior samples when 
using a population size of 150, thinning factor of 100, and burn-in period of 2000 iterations, where 
the thinning factor and burn-in period were chosen by studying the parameter chain populations and the 
average parameter autocorrelation lengths.

\subsection{Log posterior and likelihoods}
\label{sec:ext.log_posterior}

The log posterior for the combined broadband dataset given a parameter vector \pvec is
\begin{align}
 \ln P(\pvec|D) &= \ln P(\pvec) \nonumber \\
	        &+ \llh(D_\mathrm{T13}|\pvec) + \llh(D_\mathrm{M14}|\pvec) + 
\llh(D_\mathrm{F14}|\pvec) \\
		&+ \sum_i^{n_\mathrm{pb}} \llh(\Dld_{,i}|\pvec), \nonumber
\end{align}
where the first term is the log prior, followed by the per-dataset log likelihoods, and the last 
term is the sum of the \ldtk-calculated log likelihoods for the limb darkening coefficients 
for each passband.  

The exact form of the three likelihoods follows either Eq.~\eqref{eq:lnlikelihood_gn} or 
Eq.~\eqref{eq:lnlikelihood_wn}, depending on the chosen systematics model. The parametric 
(white noise) model assumes a constant average per-light-curve uncertainty (that is, the observation noise is the 
same for all the datapoints in a single light curve), and the kernels for the GP model are 
detailed below.

The light curve model consists of a product of a baseline function and a transit model 
with quadratic limb darkening calculated using \pytransit. The residual vector for a single 
light curve in Eq.~\eqref{eq:lnlikelihood_gn} is
\begin{equation}
 \vec{r} = \vec{f}_o - \vec{f}_m(\vec{X}, \pvec) = \vec{f}_o - \mathcal{B}(\vec{X}, \pvec)  \,
\mathcal{T}(\vec{t}, \pvec)
\end{equation}
where $\vec{f}_o$ is the observed flux, $\vec{f}_m$ the modelled flux, $\mathcal{B}$ the baseline 
model, $\mathcal{T}$ the transit model, $\vec{X}$ is a matrix containing the the input parameters 
(covariates), $\vec{t}$ are the mid-exposure times, and \pvec is the model parameter vector.

\subsection{Gaussian process kernels}
\label{sec:ext.gp_kernels}

The T13 and M14 datasets do not include other auxiliary information than the mid-exposure 
times. Thus, we model the noise as a Gaussian process with time as the only covariate, 
the kernel being a sum of an exponential kernel and a white noise term
\begin{equation}
k = A_t \exp\left[ - \eta_t (t_i-t_j) \right] + \sigma^2 \delta_{ij}, 
\end{equation}
where $t$ are the time values, $A_t$ is the GP output amplitude, $\eta_t$ the inverse time 
scale, $\sigma$ the average white noise standard deviation, and $\delta$ the Kronecker delta.

The F14 dataset includes also airmass, and per-observation x- and y-centroid estimates. 
This allows us to use a slightly more complex kernel, where we have an exponential time 
component, squared exponential airmass component, squared exponential PSF-centroid component, and a 
white noise term. The kernel becomes
\begin{align}
k &=  A_t    \exp\left[-\eta_t (t_i-t_j) \right] \nonumber \\
  & + A_a    \exp\left[-\eta_a (a_i-a_j)^2 \right] \nonumber \\ 
  & + A_{xy} \exp\left[-\eta_x (x_i-x_j)^2 -\eta_y (y_i-y_j)^2 \right] \nonumber\\
  & + \sigma^2\delta_{ij}, 
\end{align}
where $t$, $a$, $x$, and $y$ are the time, airmass, x, and y estimates, respectively; $A_t$, $A_a$, 
$A_{xy}$ are the GP time, airmass and xy output amplitudes; and $\eta_t$, $\eta_a$, $\eta_x$, and
$\eta_y$ the time, airmass, x and y inverse time scales.

\subsection{Results}
\label{sec:ext.results}

\begin{figure}
	\centering
	\includegraphics[width=\columnwidth]{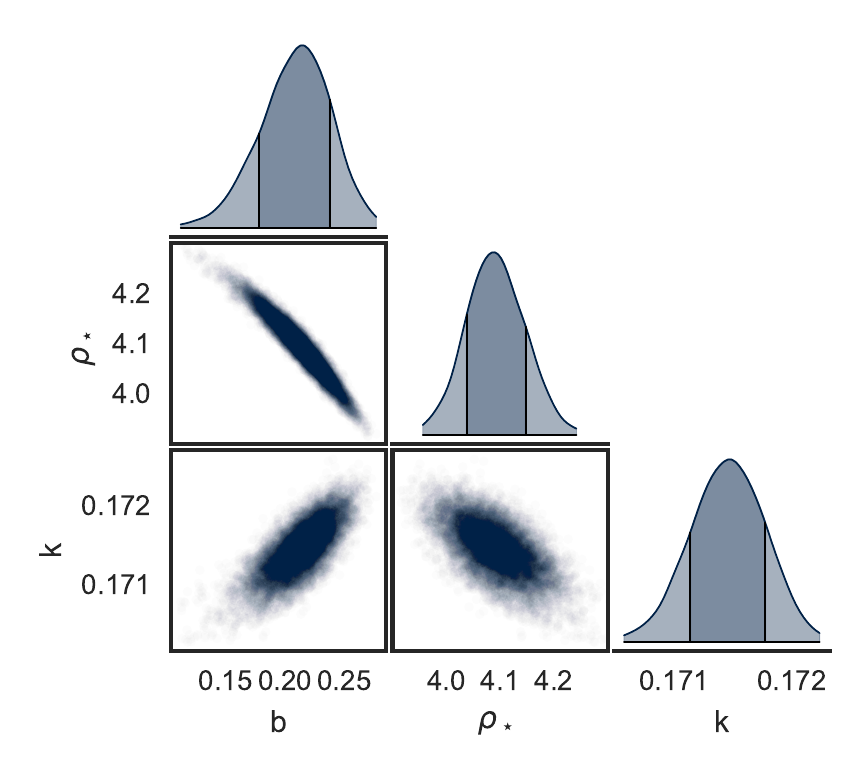}
	\caption{ \changed{Marginal- and joint-posteriors for the radius ratio, stellar density, and impact 
			parameter corresponding to the final \texttt{ckrn\textunderscore{}ldtk} run. The 68\% central
			interval is marked with a darker shade.}}
	\label{fig:ext_correlations}
\end{figure}

\begin{figure}
 \centering
 \includegraphics[width=\columnwidth]{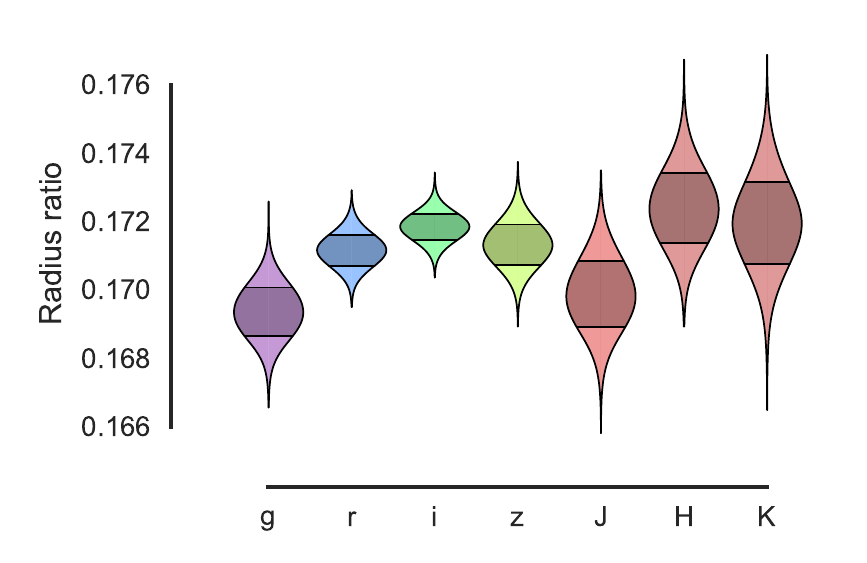}
 \caption{Broadband radius ratio posteriors estimated by jointly modelling the three prior 
datasets. The results correspond to the final  \texttt{vkrn\textunderscore{}ldtk} run with red 
noise modelled using GPs and limb darkening constrained using LDTk.}
 \label{fig:ext_radius_ratios}
\end{figure}

We adopt the \texttt{ckrn\_ldtk} (passband-independent radius ratio, GP systematics, and limb
darkening constrained using the \ldtk) run as our final analysis, and report the stellar, orbital, and 
planetary parameter estimates in Tables~\ref{tbl:ext_results} and~\ref{tbl:ext_ks}. The posterior 
distributions are all close-to normal, as shown in Figs.~ \ref{fig:ext_correlations}~and~\ref{fig:ext_radius_ratios}. 
\changed{The parameter estimates agree with the previous WASP-80b studies.}

\changed{The broadband transmission spectrum from the \texttt{vkrn\_ldtk} run, shown in Fig.~\ref{fig:ext_radius_ratios} 
is consistent with a flat line. The result agrees with a previous broadband transmission spectrum analysis by \citet{Triaud2015}. 
We used the \textsc{ExoTransmit} transmission spectrum modelling
package to test different atmosphere scenarios, but the precision in radius ratios is not sufficient 
to meaningfully distinguish between any of the physically plausible scenarios.
}

\begin{figure}
 \centering
 \includegraphics[width=\columnwidth]{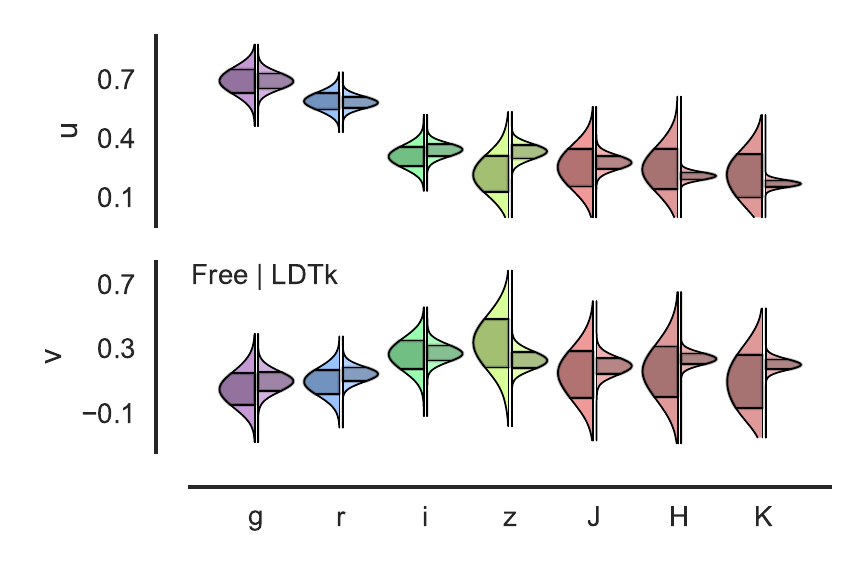}
 \caption{Broadband quadratic limb darkening coefficient posteriors. The results correspond to the 
   final \changed{\texttt{ckrn}} runs with red noise modelled with GPs and limb darkening either 
   unconstrained (left) or constrained using \textsc{LDTk} (right).}
 \label{fig:ext_limb_darkening}
\end{figure}

\begin{figure*}
 \centering
 \includegraphics[width=\textwidth]{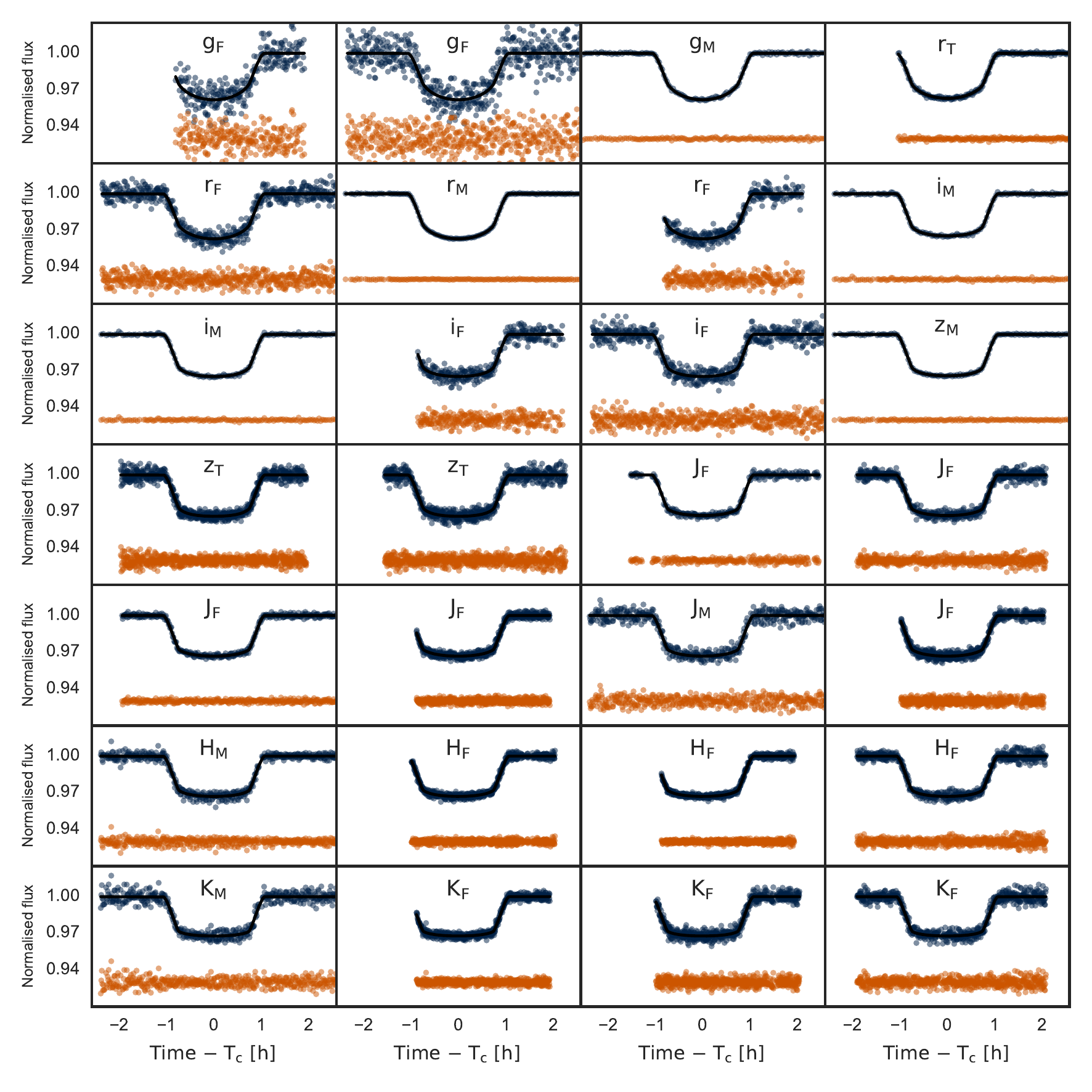}
 \caption{The 27 broadband light curves observed by \citet[][T]{Triaud2013}, 
\citet[][M]{Mancini2014}, and \citet[][F]{Fukui2014a} after the removal of the mean GP trend, the 
  posterior model medians, and the residuals.}
 \label{fig:ext_modelled}
\end{figure*}

The limb darkening coefficient posteriors are plotted in Fig.~\ref{fig:ext_limb_darkening}, both
with and without constraints from \textsc{LDTk}. The two versions agree with each other within
uncertainties. The observed light curve, conditional model distribution (for the red noise model), 
and the residuals are shown in Fig.~\ref{fig:ext_modelled}.

\section{GTC observations and data reduction} 
\label{sec:data}
\subsection{Observations}
\label{sec:data.observations}

We observed two WASP-80b transits simultaneously with one comparison star using the OSIRIS 
(Optical System for Imaging and low-Intermediate-Resolution Integrated Spectroscopy) spectrograph 
installed in the GTC (Gran Telescopio CANARIAS) on the nights starting 16~July~2014 and 
25~August~2014 (observing runs R1 and R2 respectively). The R1 observations carried from 23:30~UT 
till 3:41~UT and the R2 observations from 20:35~UT till 1:00~UT. The observing conditions were good, 
with a seeing of $\sim$0.8\arcsec{} in the beginning of each night, and all the observations were 
carried out with an airmass smaller than~1.4. 

OSIRIS \citep{Cepa1998} contains two Marconi CCD42-82 2048$\times$4096 pixel CCDs, which were used 
in the standard 2$\times$2 binning mode yielding a  plate scale of 0.254\arcsec. The observations 
were carried out using grism R1000R with a 40\arcsec-wide custom-designed slit that aims to 
minimise the systematics related to variations in flux loss. 

We chose \object{TYC~5165-00235-1} as the comparison star, located at a distance of 6.9\arcmin{} 
from WASP-80. The star has a similar V magnitude ($V=11.62$), but is slightly redder ($J=8.376$). 
The two stars were positioned equidistantly from the optical axis close to the centre of each CCD. 
The slit does not include other stars bright enough to be useful in the analysis. 

The exposure time was 6~s for R1 and 5~s for R2. While the exposure time was short, it was long 
enough for the comparison star to saturate during the final parts of the WASP-80b ingress during R1. 
This was fixed by defocusing the instrument, but meant that the comparison star cannot be used in 
the reduction during this saturated period.

For R1, 81 flat fields were taken before the transit, and 66 bias frames after the transit. For R2, 
100 flat fields were taken after the transit, and 15 bias frames before the transit. Three arc 
frames (Xe, HgAr, and Ne) for R1 were observed on 14~July and for R2 on the same night as the 
observations, after the transit.

\subsection{Data reduction and passband sets}
\label{sec:data.reduction}

\begin{figure}
 \centering
 \includegraphics[width=\columnwidth]{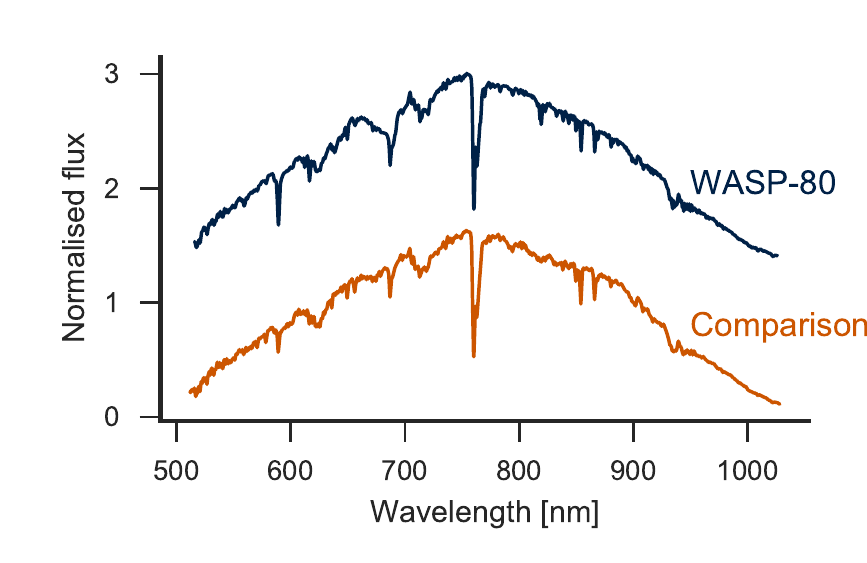}
 \caption{Normalised, sky-subtracted, and wavelength-calibrated spectra for WASP-80b (dark blue 
line) and the comparison star (orange line). }
 \label{fig:spectra}
\end{figure}

\begin{figure}
 \centering
 \includegraphics[width=\columnwidth]{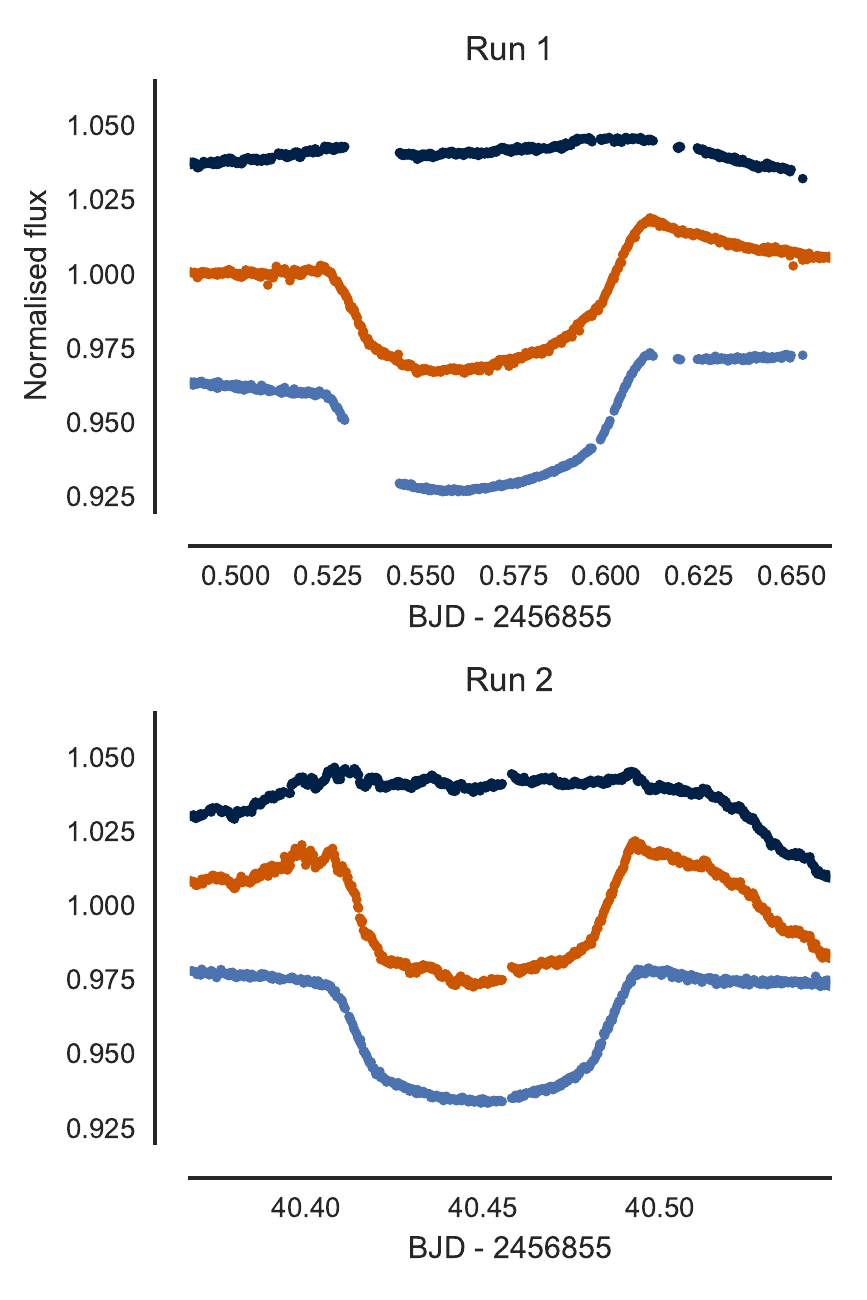}
 \caption{White-light light curves for the comparison star (top), WASP-80 (middle), and the 
relative light curve (bottom) for both observing runs. A small number of clear outliers have been 
omitted for clarity.}
 \label{fig:white_lcs}
\end{figure}

The passband-integrated light curves are produced from the raw data using the pipeline described in \citet{Chen2016,Chen2017}. 
The pipeline carries out the basic CCD data reduction steps, calculates a 2D wavelength solution, removes the sky, and
generates a reduced 1D spectrum for WASP-80 and the comparison star (Fig.~\ref{fig:spectra}). The light curves are then 
generated by integrating the spectra multiplied by a transmission function defining the passband. We created four 
sets of passbands (datasets) listed in Table~\ref{tbl:gtc_datasets}. \changed{The \dsw
dataset consists of a single light curve covering the whole usable spectrum (white light curve), the \dsnb (narrow-band dataset) covers the whole
usable spectrum in 20~nm-wide bins, and the \dsna and \dsk datasets cover the Na and K lines, respectively, in 6~nm-wide bins.}
The final white-light WASP-80 and reference star lightcurves with the relative light curve are shown in Fig.~\ref{fig:white_lcs}. 

\changed{The white-light white noise estimates for R1 and R2 are 400 and 520~ppm, respectively. For the \dsnb
narrow-band dataset, the white noise level varies from 600 to 2100~ppm, with a median level of  720 and 870~ppm
for R1 and R2, respectively. The difference in the white noise levels is expected due to the different exposure times
used in R1 and R2 (6 and 5~s, respectively).
}

\begin{table}[t]    
	\caption{Passband sets extracted from the GTC spectroscopy.}
	\centering
	\begin{tabular*}{\columnwidth}{@{\extracolsep{\fill}} llll}
		\toprule\toprule
		Dataset name & $N_{pb}$ & Width [nm] & Range [nm]\\
		\midrule
		\changed{\dsw}    & 1 &   375  & 520---900 \\
		\changed{\dsnb}  & 19 & 20  & 520---900 \\
		\changed{\dsk}     & 9 &  6 & 737---795\\
		\changed{\dsna}  & 9 &   6 &  562---616\\
		\bottomrule
	\end{tabular*}
	\label{tbl:gtc_datasets}
\end{table}

\subsection{Systematics}
\label{sec:observations.systematics}

The relative light curves in Fig.~\ref{fig:white_lcs} feature systematics not corrected by division 
by the comparison star. Specially, the baseline is affected by 
a smooth trend that cannot be modelled by a simple parametric model as a function of time (or any 
of the simultaneously measured auxiliary variables). The trend causes a pre- and post-transit 
baseline difference of $\sim2\%$ during run 1, and a $\sim1\%$ difference during run~2. A similar, 
smooth, nearly linear, variation as a function of the rotator angle accompanied by relatively 
smooth 'bumps', has been observed in previous OSIRIS transmission spectroscopy studies, and is 
likely caused by vignetting in the telescope pupil space \citep{Nortmann2016}. 

Fortunately, the systematics are mainly common-mode, without significant variation across
the spectrum. Common-mode systematics can be removed by either fitting the white-light light 
curves with a flexible GP, which can be used to create a common-mode systematics model, or 
by using a divide-by-white approach. After the common-mode correction, the residual 
wavelength-dependent systematics can be accounted for either with parametric or non-parametric 
approaches, both of which were used in our analyses.

\section{GTC analysis}
\label{sec:gtc_analysis}

\subsection{Overview}
\label{sec:gtc_analysis.overview}

The GTC transmission spectroscopy covers the modelling of the white light curves and the three 
narrowband datasets described  in Sec.~\ref{sec:data.reduction}. We used three
modelling approaches:
\begin{description}
	\item[\mdr] direct modelling with flexible GP-based systematics
	\item[\changed{\mww}] divide-by-white with parametric systematics
	\item[\mwr] divide-by-white with GP-based systematics
\end{description}
The first approach, direct modelling of the light curves with a flexible GP-based systematics model,
was used to obtain a model for the common-mode systematics (from the white light curves), and to
improve the system characterisation (from the full-spectrum 20~nm-wide dataset). The divide-by-white (DW) approaches
were then used for transmission spectroscopy, with the direct-modelling posteriors used as priors on the
wavelength-independent system parameters (the motivation for this is discussed later). The analyses 
were repeated with and without \ldtk to assess how sensitive the results are to assumptions about limb 
darkening, and modelling the two nights separately and jointly, to test whether the results are consistent 
from night to night. 

The parameter posteriors are estimated in similar fashion to the prior data analysis in 
Sec.~\ref{sec:ext.overview}. Differential evolution is used to create a parameter vector population
clumped around the global posterior maximum, which used to initialise the \textsc{emcee} sampler.
The sampler is  ran over $10 
\times 15\;000$ iterations (that is, ten sets of $15\;000$ iterations where each set is initialised 
from the final state of the previous set) with 600 chains and a thinning factor of 100, which 
gives us a final set of 75$\;$000 independent posterior samples. The number of iterations, chains,
and the thinning factor were chosen after studying the chain populations and per-parameter
autocorrelation lengths.

Unlike in the broadband dataset analysis, we keep the GP hyperparameters free and marginalize over them. 
This slows down the sampling process, but yields more reliable posteriors, and allows us to study how 
well the GP kernels represent the systematics.

\subsection{Modelling approaches}
\subsubsection{Direct model with GP systematics}

The direct-modelling approach \mdr reproduces the light curves as a product of a baseline and a transit 
model directly. We model all the passbands for both nights in a dataset 
jointly, which leads to a slightly involved parameterisation, but aims to utilise the data fully. As mentioned
earlier, the model 
parameters can be divided into four categories: a) passband- and baseline-independent parameters, 
b) achromatic per-night baseline parameters, c) chromatic per-light-curve parameters, and d) chromatic 
parameters that should stay constant between observation runs. 

The direct model parametrisation and the parameter priors are listed in Table~\ref{tbl:dm_parameters}. 
Priors for the zero epoch, orbital period, impact parameter, and stellar density are based on the broadband dataset 
analysis posteriors (run \texttt{ckrn\_ldtk}). The $u$ and $v$ limb darkening coefficients correspond 
to the \citet{Kipping2013b} quadratic limb darkening model parametrisation, where uniform priors 
from 0 to 1 lead to uninformative priors covering the physically viable values for the quadratic limb darkening priors.

The GP hyperparameters can be chosen to be night- or light-curve-dependent, but we choose to
keep them independent for practical reasons. Our approach adds three GP hyperparameters to the analysis,
and requires two covariance matrix inversions per posterior evaluation (one per night). Making the 
GP hyperparameters light-curve-dependent would add three free parameters to the model for each light curve,
and require a covariance matrix inversion for each light curve, which would make marginalisation over
the GP hyperparameters costly. Making the GP hyperparameters night-dependent would be feasible, since
the approach would add only three more parameters to the model, and would not require additional covariance
matrix inversions. However, our analyses for separate nights result with compatible GP hyperparameter 
posteriors, and we choose the simplest approach.

\begin{table}[t]
 \caption{Model parametrisations and priors. \UP{a,b} stands for a uniform prior from a  to b, where a and
 	b are omitted when the range is chosen to be wide enough not to affect the posteriors. \NP{\mu,\sigma} 
 	stands for a normal prior with mean $\mu$ and standard deviation $\sigma$. The system parameters
    have normal priors with means and standard deviations corresponding to the values shown in Table~\ref{tbl:ext_results}.}
 \centering
 \begin{tabular*}{\columnwidth}{@{\extracolsep{\fill}}lll}
  \toprule\toprule
  Notation & Name & Prior \\
  \midrule
  \multicolumn{3}{l}{\textit{System parameters, passband independent}} \\
  $T_c$  & transit centre & $\mathcal{N} $\\
  $P$    & orbital period & $\mathcal{N} $ \\
  $\rho_\star$ & stellar density & $\mathcal{N} $ \\
  $b$ & impact parameter & $\mathcal{N} $ \\
  \\
  \multicolumn{3}{l}{\textit{Uninformative priors, passband dependent}} \\
  $k^2$ & area ratio & $\UP{0.165^2, 0.175^2}$ \\
  $q_1$ & limb darkening q$_1$ & $\UP{0, 1}$ \\
  $q_2$ & limb darkening q$_2$ & $\UP{0, 1}$ \\
  \\
  \multicolumn{3}{l}{\textit{Uninformative priors, light curve dependent}} \\
  $c_b$ & baseline constant & $\UP{-1, 1}$ \\
  $c_t$ & linear time coefficient\tablefootmark{a} & $\UP{-1, 1}$ \\
  $c_x$ & linear airmass coefficient\tablefootmark{a} & $\UP{-1, 1}$ \\
  \\
\multicolumn{3}{l}{\textit{GP hyperparameters\tablefootmark{b}, night dependent}} \\
$\log \gamma_x$ & log GP airmass scale & $\UP{-5, -1}$ \\
$\log a_\alpha$ & log GP rotator angle amplitude & $\UP{-5, -1}$ \\
$\log \gamma_\alpha$ & log GP rotator angle input scale & $\UP{-5, \hphantom{-}3}$ \\    
  \bottomrule
 \end{tabular*}
\tablefoot{
	\tablefoottext{a}{Only for the parametric systematics model.}
	\tablefoottext{b}{Only for the nonparametric systematics model.}}
 \label{tbl:dm_parameters}
\end{table}

The direct model represents each light curve as a Gaussian process
\begin{equation}
	f \sim \mathcal{N}(c_b\;\mathcal{T}(\vec{t}, \pvec), \covmat),
\end{equation}
where $c_b$ is a baseline constant, $\mathcal{T}$ is the transit model, and \covmat
is the covariance matrix. We use airmass $x$, and telescope rotator angle, $\alpha$, 
as GP input parameters (covariates), with a covariance matrix defined as a sum of a
linear kernel and  squared-exponential kernel,
\begin{equation}
	\covmat(i,j) = \left( \frac{x_j \cdot x_i}{\gamma_x}\right) + a_\alpha^2 \exp \left( \frac{(\alpha_j - \alpha_i)^2}{\gamma_\alpha}\right) + \delta_{ij} \sigma
\end{equation}
where $a_\alpha$ is an output scale parameter, $\gamma_x$ and $\gamma_\alpha$ are the 
input scale parameters, and $\sigma$ is the average white noise. Any residual airmass-dependent systematics 
should be approximately linear, and can be modelled with a linear kernel, while the possible chromatic rotator-angle 
dependencies are expected to be smooth, and  well-modelled by a squared-exponential kernel.

We also tested a GP with time as a covariate (with a Matern kernel), but it did not affect the posteriors significantly. Also,
the power spectral density (PSD) of the residuals with the transit and the two-covariate GP mean removed is
approximately constant, which suggests that the noise is white after the residual airmass and rotator-angle 
dependencies are accounted for.

The white noise level does not vary significantly from passband to passband in our datasets, and we choose to
use a single average white noise estimate for an observing run (again, so that we do not need to invert the covariance
matrix separately for each light curve). The white noise estimate is an average of the estimates calculated separately 
for each light curve, which are calculated as 
\begin{equation}
\sigma= \mathrm{std}(\vec{f}_d) / \sqrt{2}
\end{equation} where
 $\vec{f}_d = f_i - f_{i-1}$. That is, we estimate the white noise from the standard deviation of the light curve 
 differentials.

The radius ratios and limb darkening coefficients yield three free parameters per passband, and the baseline constant 
a further free parameter per light curve. In total, the number of free parameters for a direct model reaches 108 for 
the 20-nm run with 20 passbands. The computations can still be carried out with a modern laptop in a matter of 
hours\footnote{\pytransit, which was used to compute the light curves, is optimised to compute a set of 
multiband light curves with only a minor additional cost to calculating a single passband.} for a single 
analysis\footnote{However, while a single analysis can be carried out in a laptop, the Glamdring-cluster in Oxford 	
University and the TeideHPC supercomputer in Spain were used to carry out the final analyses.}, but tests must be carried
out to ensure that the final posterior sample gives a reliable representation of the true posterior distribution.

\subsubsection{Divide-by-white model with and without GP systematics}

The divide-by-white (DW) approach\footnote{Our approach would be more accurately named as divide-by-average,
since we're dividing each dataset by the dataset mean.} allows us to remove any common-mode systematics from
the dataset with the cost of increased per-passband white noise. The residual vector $\vec{r}$ in 
Eq.~\eqref{eq:lnlikelihood_gn} for the DW approach is
\begin{equation}
\vec{r} = \frac{1}{\vec{b}(\pvec, \mathbf{X})} \left( \frac{\vec{f}}{1/\npb \sum 
\vec{f}} - \frac{\tmodel(\vec{t}, \pvec)}{1/\npb \sum \tmodel(\vec{t}, \pvec )} \right ), \label{eq:dw_residuals}
\end{equation}
where \flux is the observed flux vector, \tmodel is the transit model, and $\vec{b}$ is 
the residual baseline model. That is, we divide both the observed and modelled fluxes by the values 
averaged over all the passbands in the dataset.

The DW approach does not only remove the common-mode systematics, but all colour-independent
signals. Effectively, the signals left in the data are due to colour-dependent systematics, changes in the
radius radio, and changes in stellar limb darkening. The system parameters, such as the orbital period
and impact parameter, are poorly constrained, so we set informative priors based 
on the \texttt{nb\_ldtk} direct-model analysis on them. The average radius ratio is not well
constrained either, so we set a prior on the median radius ratio based on the broadband dataset analysis. 
We use median instead of mean to ensure that the prior does not affect the scaling of the transmission spectrum, as using mean would.

We repeat the DW analysis using first a parametric baseline model (\mww approach), and then a more flexible
Gaussian process-based baseline model (\mwr approach). The parametric model represents  the baseline for 
each light curve as a linear function of time and airmass
\begin{equation}
  \vec{b}_i(\vec{t}, \vec{x}, \pvec ) = \theta_{i,a} + \theta_{i,b} \vec{t} + \theta_{i,c} \vec{x},
\end{equation}
where $\theta_{i,a}$, $\theta_{i,b}$, and $\theta_{i,b}$ are the light curve specific baseline 
constant, linear time coefficient and linear airmass coefficient, respectively.  The residual noise
is considered white, which makes the approach significantly faster than using GPs, but yields
three free parameters per light curve to the model.

The \mwr analysis uses the same GP kernel as the \mdr approach, changing only the GP mean 
function to the one in Eq.~\ref{eq:dw_residuals}.

\section{Results}
\label{sec:results}

\subsection{System characterisation}
\label{sec:results.broadband}

The direct-modelling \mdr runs were used to improve the system characterisation from the broadband dataset analysis,
and yield the final WASP-80b parameter estimates listed in Table~\ref{tbl:system_characterisation}. 
The estimates correspond to the full-spectrum narrow-band dataset (\dsnb) with \ldtk. The narrow-band 
run was preferred over the white-light run since the colour information allows us to mitigate the 
degeneracy between  the average impact parameter, radius ratio, and limb darkening, leading to 
improved posterior estimates.

\begin{table}[t]    
	\caption{Results from the final GTC characterisation run (\texttt{nb\textunderscore{}12\textunderscore{}ldtk}
		with systematics modelled using GPs, wavelength-dependent radius ratio, and limb darkening
		constrained with \textsc{LDTk}). The uncertainty is based on the central 68\% posterior 
		interval. We do not list the radius ratio estimates, since the (common-mode)
		systematics are too strong for direct modelling to constrain them.}
	\centering
	\begin{tabular*}{\columnwidth}{@{\extracolsep{\fill}} llll}
		\toprule\toprule
		Parameter & Units & Posterior median & Uncertainty \\
		\midrule
		Zero epoch       & BJD& 2456125.41737  & 8.4e-05 \\ 
		Period           &days&       3.067855 & 3.6e-07 \\
		Impact parameter &    &       0.161    & 1.7e-02 \\
		Stellar density  &\gcm&       4.172    & 3.3e-02 \\
		\bottomrule
	\end{tabular*}
	\label{tbl:system_characterisation}
\end{table}

\subsection{Transmission spectroscopy}
\label{sec:results.transmisison_spectroscopy}

\begin{figure*}
	\centering
	\includegraphics[width=\textwidth]{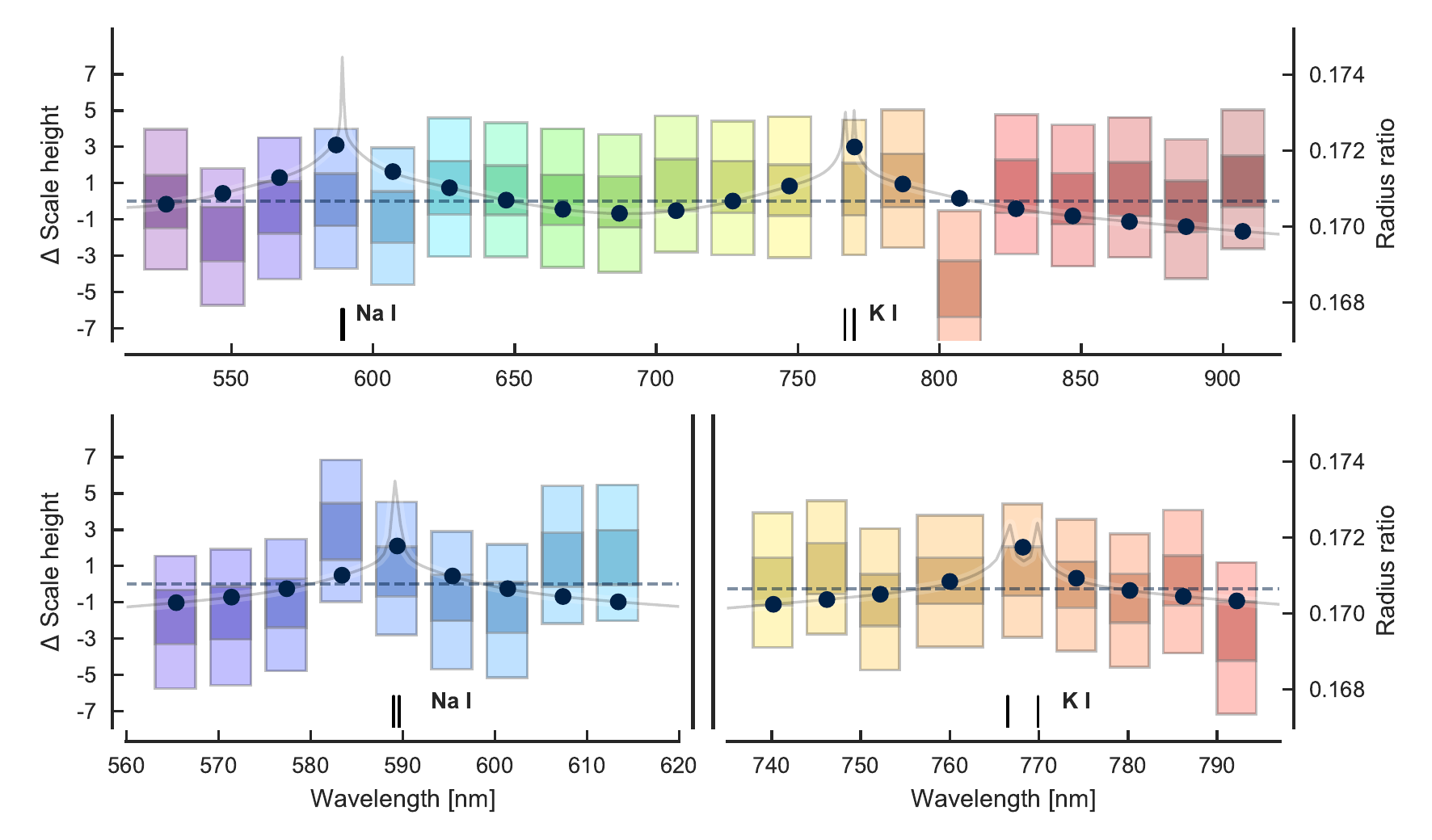}
	\caption{Transmission spectra for the \dsnb dataset with 20 passbands spanning from 520~nm to 920~nm (top),
		the \dsna set covering the Na~I doublet (bottom left), and the \dsk dataset covering the K~I double (bottom right). 
		The inner boxes correspond to the central 68\% posterior intervals, the outer boxes to the central 99\% posterior 
		intervals, and Na~I and K~I lines are marked as black lines. The light gray line shows an \textsc{Exo-Transmit} spectrum
		with Na and K described in Sect.~\ref{sec:discussion.comparison}, the black dots show the \textsc{Exo-Transmit} 
		spectrum binned to the dataset binning, and the slashed horizontal line shows the transmission spectrum mean. }
	\label{fig:ts_nb}
\end{figure*}

We show the transmission spectra for the \dsnb, \dsk, and \dsna datasets
in Fig.~\ref{fig:ts_nb}. These results correspond to the joint \mwr run with a GP systematics model and limb 
darkening coefficients constrainted by \ldtk. The figures show the central 68\% and 99\% radius ratio
posterior intervals, and the point estimates are listed in Table~\ref{tbl:transmission_spectra}.

\begin{table}[t]    
	\caption{Passband centres, ranges, and radius ratio estimates for the three narrow-band datasets
		\changed{from the final \mwr approach.}}
	\centering
	\begin{tabular*}{\columnwidth}{@{\extracolsep{\fill}} ccl}
		\toprule\toprule
		Pb. centre [nm] & Pb. range [nm] & Radius ratio \\
		\\
		\multicolumn{3}{l}{Full-spectrum dataset}\\
		\midrule
		527 & 517---537 &0.1707 $\pm$ 0.00070 \\
		547 & 537---557 &0.1698 $\pm$ 0.00071 \\
		567 & 557---577 &0.1705 $\pm$ 0.00068 \\
		587 & 577---597 &0.1707 $\pm$ 0.00069 \\
		607 & 597---617 &0.1703 $\pm$ 0.00067 \\
		627 & 617---637 &0.1710 $\pm$ 0.00070 \\
		647 & 637---657 &0.1709 $\pm$ 0.00065 \\
		667 & 657---677 &0.1707 $\pm$ 0.00066 \\
		687 & 677---697 &0.1707 $\pm$ 0.00067 \\
		707 & 697---717 &0.1711 $\pm$ 0.00068 \\
		727 & 717---737 &0.1710 $\pm$ 0.00068 \\
		747 & 737---757 &0.1709 $\pm$ 0.00067 \\
		770 & 764---775 &0.1709 $\pm$ 0.00069 \\
		787 & 777---797 &0.1712 $\pm$ 0.00070 \\
		807 & 797---817 &0.1683 $\pm$ 0.00074 \\
		827 & 817---837 &0.1710 $\pm$ 0.00070 \\
		847 & 837---857 &0.1707 $\pm$ 0.00067 \\
		867 & 857---877 &0.1710 $\pm$ 0.00071 \\
		887 & 877---897 &0.1705 $\pm$ 0.00068 \\
		907 & 897---917 &0.1712 $\pm$ 0.00067 \\
		\\
		\multicolumn{3}{l}{Na~I dataset}\\
		\midrule
		565 & 562---568 &0.1699 $\pm$ 0.00071 \\
		571 & 568---574 &0.1701 $\pm$ 0.00070 \\
		577 & 574---580 &0.1703 $\pm$ 0.00064 \\
		583 & 580---586 &0.1722 $\pm$ 0.00074 \\
		589 & 586---592 &0.1710 $\pm$ 0.00065 \\
		595 & 592---598 &0.1705 $\pm$ 0.00061 \\
		601 & 598---604 &0.1702 $\pm$ 0.00067 \\
		607 & 604---610 &0.1714 $\pm$ 0.00071 \\
		613 & 610---616 &0.1715 $\pm$ 0.00072 \\
		\\
		\multicolumn{3}{l}{K~I dataset}\\
		\midrule
		740 & 737---743 &0.1708 $\pm$ 0.00063 \\
		746 & 743---749 &0.1711 $\pm$ 0.00067 \\
		752 & 749---755 &0.1704 $\pm$ 0.00068 \\
		760 & 755---765 &0.1708 $\pm$ 0.00061 \\
		768 & 765---771 &0.1711 $\pm$ 0.00064 \\
		774 & 771---777 &0.1707 $\pm$ 0.00061 \\
		780 & 777---783 &0.1704 $\pm$ 0.00064 \\
		786 & 783---789 &0.1708 $\pm$ 0.00066 \\
		792 & 789---795 &0.1695 $\pm$ 0.00078 \\
		\bottomrule
	\end{tabular*}
	\label{tbl:transmission_spectra}
\end{table}

Comparison of the results from the two different divide-by-white approaches can be found below in
Sect.~\ref{sec:appendix_1}, and a comparison of the results obtained using only the light curves from 
one of the two can be found Sect.~\ref{sec:appendix_2}. We do not show the results for the analysis 
without \ldtk. Unconstrained limb darkening does not have a significant effect on the radius ratio 
posterior medians, but does increase the posterior width significantly due to the strong degeneracy 
caused by the DW approach. 

\begin{figure*}
 \centering
 \includegraphics[width=\textwidth]{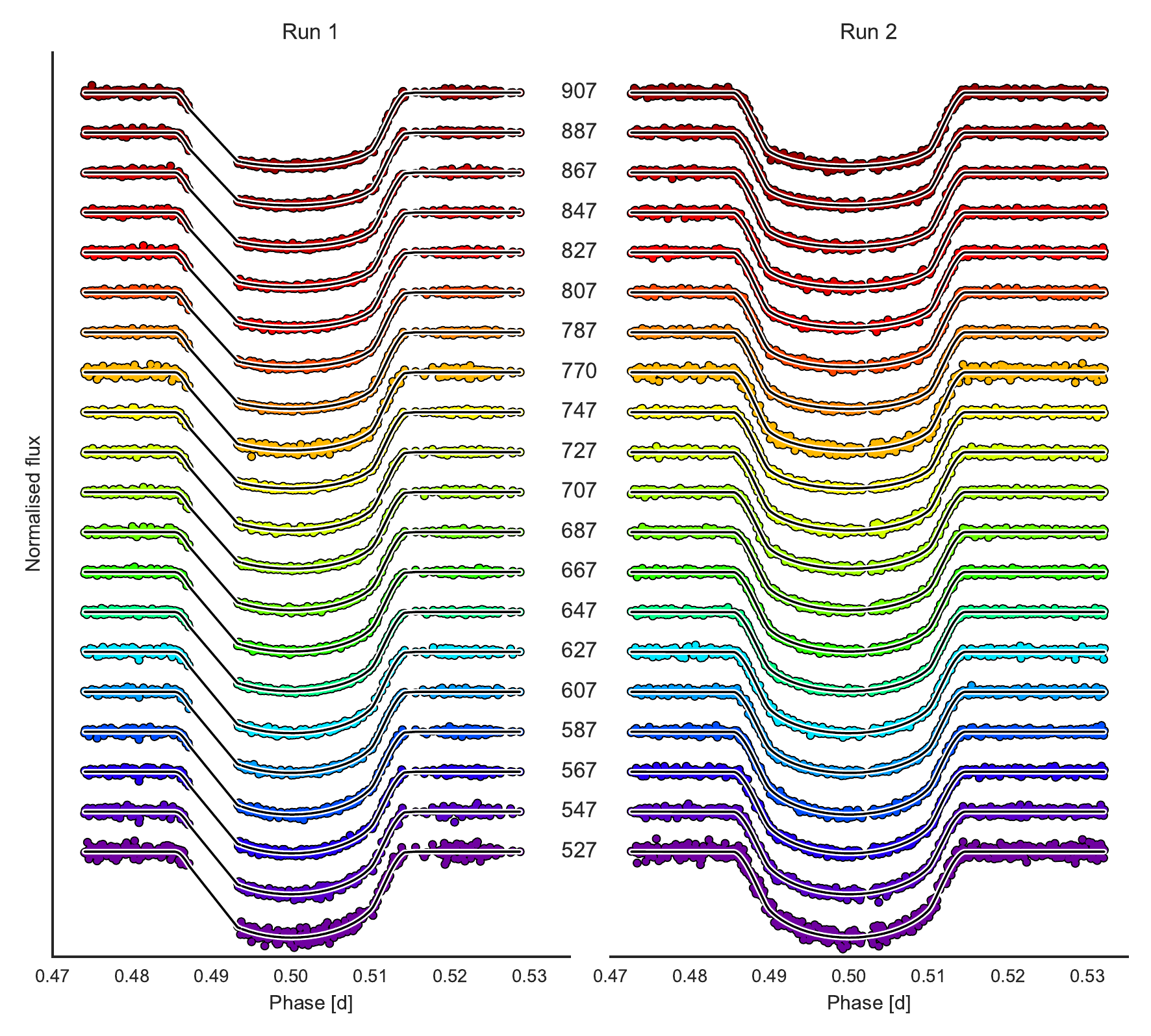}
 \caption{The \dsnb  light curves for both observing runs and the \mdr approach posterior median model  
 	(black line) as a function of the phase. Passband centres are marked in nanometres.}
 \label{fig:lcs_direct_nb}
\end{figure*}

\section{Discussion}
\label{sec:discussion}
\subsection{Flat transmission spectrum}

We adopt the results from the more conservative approach, \mwr, as the final results of the analysis.
All the spectra in Fig.~\ref{fig:ts_nb} are flat within 
uncertainties\changed{, in agreement with the broadband analysis, and  the broadband transmission spectroscopy 
carried out by \citet{Triaud2015}. } The \dsnb features a single strongly deviating
passband centred at 807~nm, with a significantly smaller radius ratio than the average. The passband
is likely affected by instrumental observation-geometry-dependent systematics (the deviation is evident
in both observing runs, albeit slightly different), and we discuss it in more detail in Sect.~\ref{sec:discussion.problem_pb}.

The flat transmission spectrum allows us to rule out strong Rayleigh scattering or Na~I or K~I absorption in 
WASP-80b's atmosphere, but does not justify detailed atmospheric modelling. However, basic modelling with 
\textsc{Exo-transmit} \citep{Kempton2017} strongly favours a flat spectrum over a spectrum from an atmosphere with 
0.1 or 1 solar metallicity and K and Na: a likelihood ratio test between the \textsc{Exo-Transmit}
spectra  and a flat spectrum favours the flat model with likelihood ratios around 1000-5000. (We fit the model spectra
to the estimated spectrum with a free scaling factor, and the flat spectrum is the mean of the estimated spectrum.)
Figure~\ref{fig:ts_nb} includes an \textsc{Exo-Transmit} spectrum with K and Na calculated assuming Solar metallicity, gas-phase 
chemistry, $T_\mathrm{eq}=800$~K, $\logg = 3.18$, $\rstar = 0.57 R_\odot$,  and $k=0.17$ as an example.
  
The \mwr approach was chosen over \mww because the \mww results show greater discrepancies in the spectra
estimated from R1 and R2 separately, as discussed below. It is likely that the parametric systematics model is not 
flexible enough, and the unaccounted-for systematics lead to biases in the radius ratio values. The GP kernel
in the \mwr approach is flexible enough to marginalize over the systematics, and still allows for a reliable radius
ratio estimation (tested with two mock datasets discussed in Sect.~\ref{sec:discussion.mock}).

\subsection{Comparison with the previous K~I detection}
\label{sec:discussion.comparison}

\begin{figure}
	\centering
	\includegraphics[width=\columnwidth]{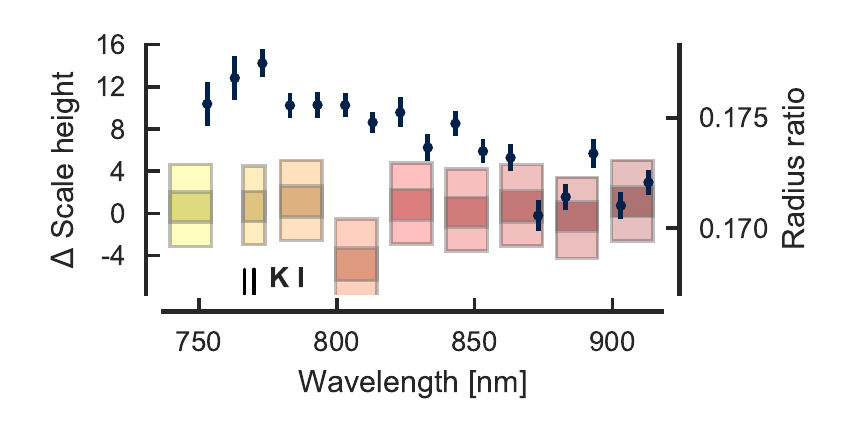}
	\caption{Comparison between the results from the GTC analysis and the results presented by \citet{Sedaghati2017}.}
	\label{fig:ts_comparison}
\end{figure}

\citet{Sedaghati2017} reported of a detection of strong K~I absorption in WASP-80b's atmosphere
based on transmission spectroscopy with the FORS2 spectrograph. Our results disagree with the reported K~I 
detection, as shown in Fig.~\ref{fig:ts_comparison}, but agree with the redmost part of their spectrum.

Both of our divide-by-white approaches are consistent with each other, as are the transmission spectra
estimated separately from the two observing runs (as discussed below in Sects.~\ref{sec:appendix_1}~and~\ref{sec:appendix_2}),
which leads us to believe that the strong K~I signal reported by \citet{Sedaghati2017} is due to systematics, even though
the analysis described in \changed{their} paper seems rigorous in all standards.

\subsection{Comparison of modelling approaches}
\label{sec:appendix_1}

The two divide-by-white approaches differ in their way of modelling the systematics. The \mww approach
models the systematics with parametric model as a combination of linear functions of time and airmass, and 
assumes that the everything else is white normally distributed noise, while the \mwr approach models the 
systematics as a Gaussian process. In the \mww case all systematics not corrected by the division by the dataset
average light curve or modelled by the parametric model may cause biases in the radius ratio estimates.
The GP kernel in the \mwr case is relatively flexible, even when we impose a linear relation on airmass,
and should yield  less bias-prone radius ratio estimates. This is true especially since we're integrating over 
the GP hyperparameters, and do not significantly constrain them with priors\footnote{The GP hyperparameters
are actually allowed to probe the parameter space where they model time correlation rather than rotator-angle
correlation, since the rotator angle itself is a slowly and smoothly varying function of time. }

\begin{figure*}
	\centering
	\includegraphics[width=\textwidth]{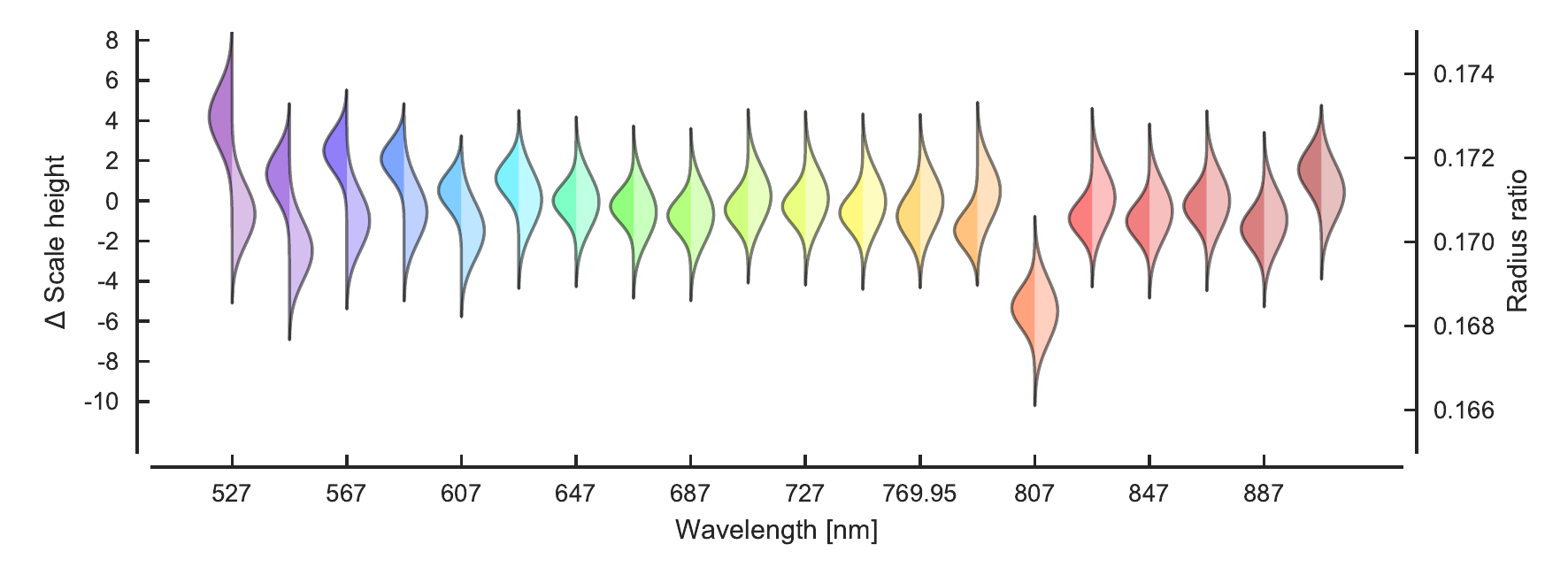}
	\caption{Radius ratio posterior distributions approximated as normal distributions for the NB DWW (left) and DWR (right) runs. }
	\label{fig:pub_ts_nb_dwdwr}
\end{figure*}

\begin{figure}
	\centering
	\includegraphics[width=\columnwidth]{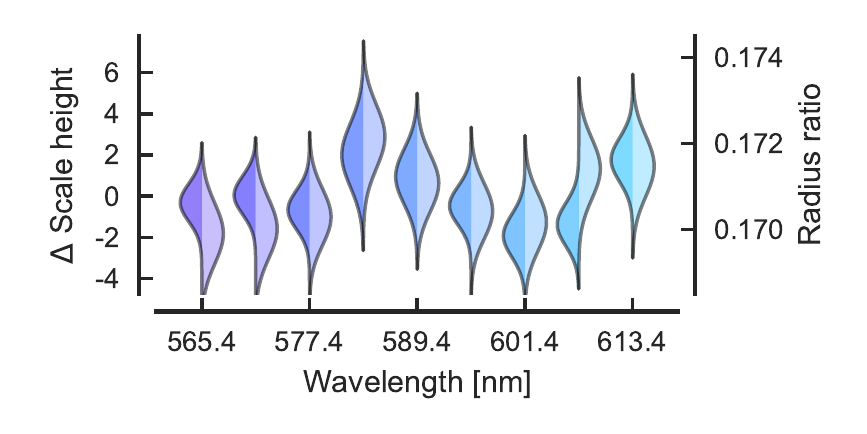}
	\caption{Radius ratio posterior distributions approximated as normal distributions for the Na DWW (left) and DWR (right) runs. }
	\label{fig:pub_ts_na_dwdwr}
\end{figure}

\begin{figure}
	\centering
	\includegraphics[width=\columnwidth]{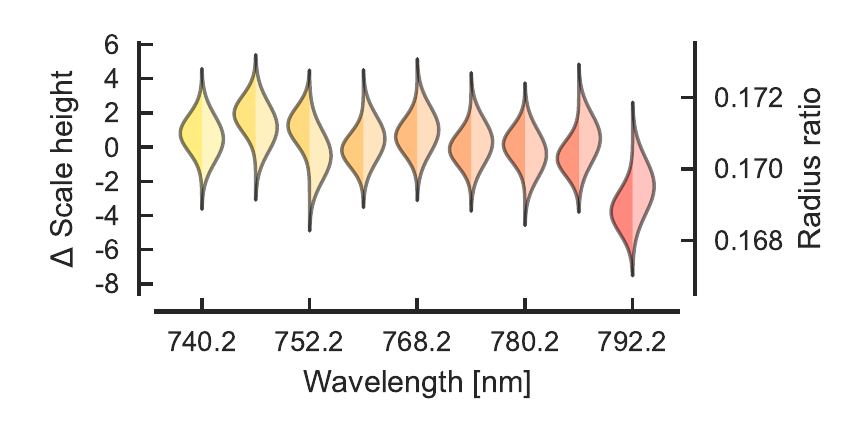}
	\caption{Radius ratio posterior distributions approximated as normal distributions for the K DWW (left) and DWR (right) runs. }
	\label{fig:pub_ts_k_dwdwr}
\end{figure}

We compare the transmission spectra obtained with \mww and \mwr approaches in Figs.~\ref{fig:pub_ts_nb_dwdwr},
\ref{fig:pub_ts_na_dwdwr}, and \ref{fig:pub_ts_k_dwdwr}. The differences are small for the Na and K analyses,
which can be expected due to the small wavelength range covered by the datasets. However, the \dsnb set 
analysed with \mww approach features a Rayleigh-like signal in the bluemost passbands that is not visible in
the \mwr results. When analysing the two nights separately (Sect.~\ref{sec:appendix_2}), we can see that the 
Rayleigh-like signal arises from the second night, and is not significant in the first night. However, the first night 
shows an increasing trend towards the red end of the spectrum, not visible in the second night spectrum.

In theory, the varying Rayleigh-like slope could be interpreted as variations in stellar activity. However, a more
likely explanation is residual wavelength-dependent systematics not accounted for properly by the parametric model.

\subsection{Comparison of individual nights}
\label{sec:appendix_2}

\begin{figure*}
	\centering
	\includegraphics[width=\textwidth]{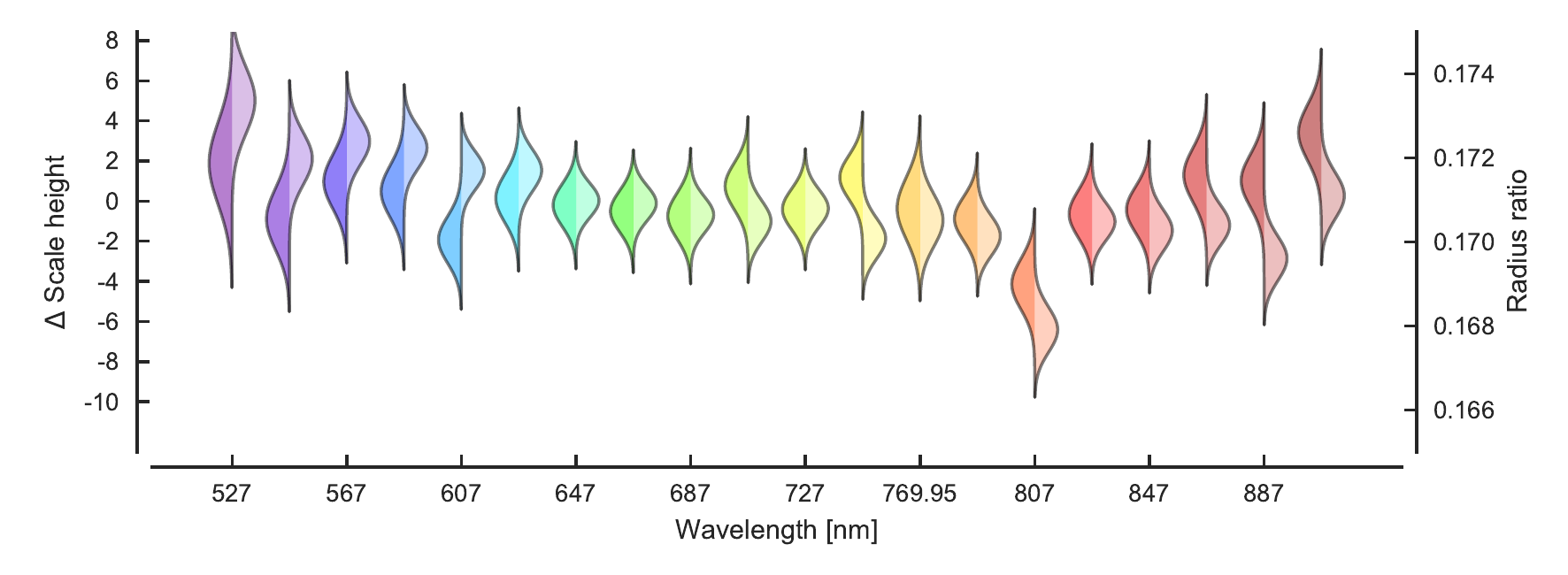}
	\includegraphics[width=\textwidth]{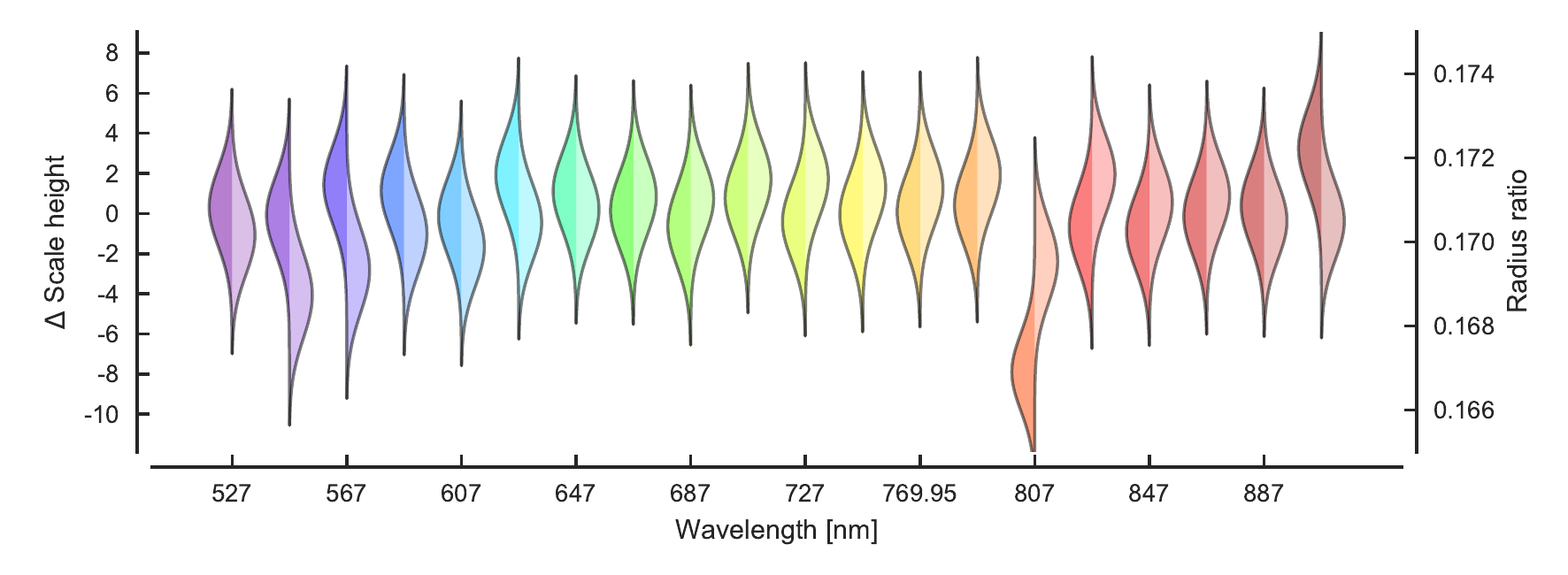}
	\caption{Radius ratio posterior distributions approximated as normal distributions for the \dsnb \mww (up) and \mwr (below) runs
		using only the data from night 1 (left) or night 2 (right). }
	\label{fig:comparison_nights_nb}
\end{figure*}

\begin{figure}
	\centering
	\includegraphics[width=\columnwidth]{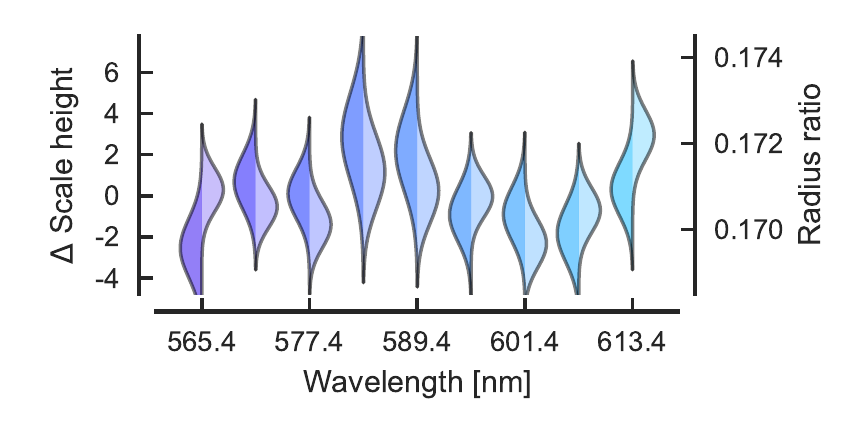}
	\includegraphics[width=\columnwidth]{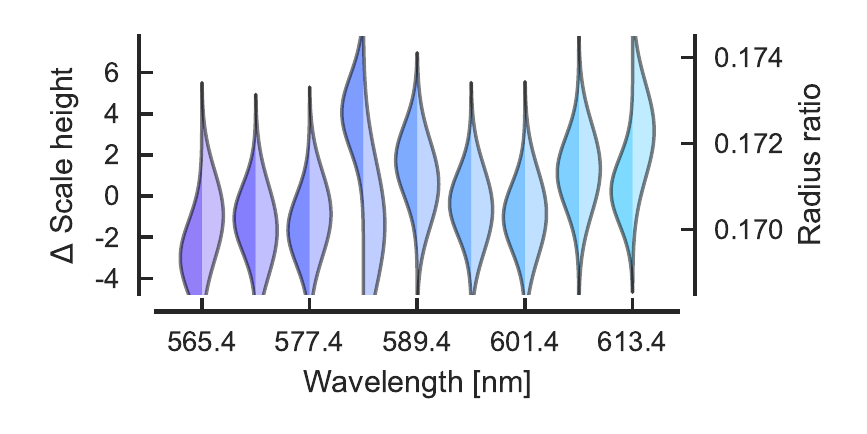}
	\caption{Radius ratio posterior distributions approximated as normal distributions for the \dsna \mww (up) and \mwr (below) runs
		using only the data from night 1 (left) or night 2 (right). }
	\label{fig:comparison_nights_na}
\end{figure}

\begin{figure}
	\centering
	\includegraphics[width=\columnwidth]{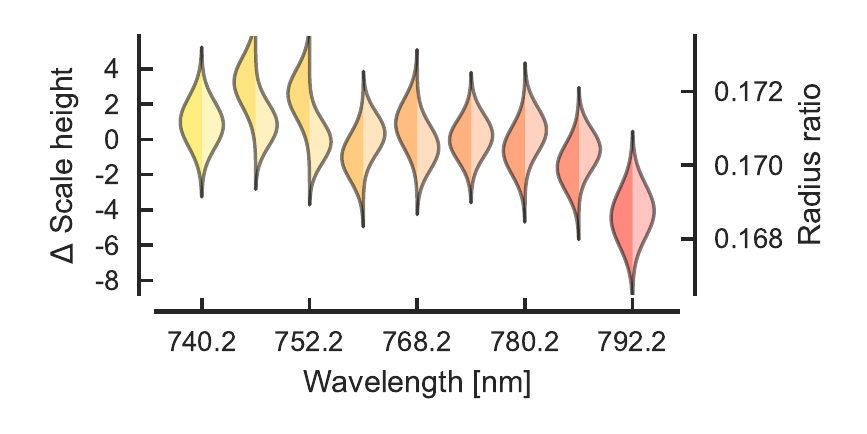}
	\includegraphics[width=\columnwidth]{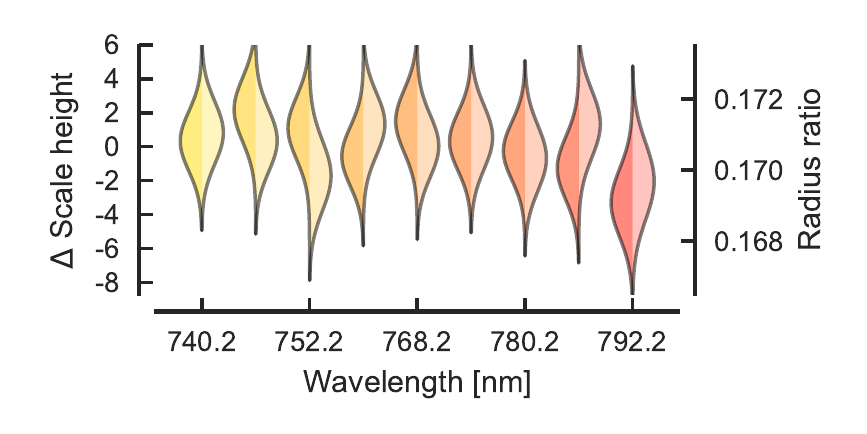}
	\caption{Radius ratio posterior distributions approximated as normal distributions for the \dsk \mww (up) and \mwr (below) runs
		using only the data from night 1 (left) or night 2 (right). }
	\label{fig:comparison_nights_k}
\end{figure}

We show the transmission spectra estimated separately from run 1 or run 2 either with the \mww or \mwr approach in 
Figs~\ref{fig:comparison_nights_nb},~\ref{fig:comparison_nights_na}, and \ref{fig:comparison_nights_k}. The results from the
different nights agree with some exceptions. Both the blue and red ends of the \mww \dsnb spectrum differ systematically, as mentioned
earlier.  the Rayleigh-like 
signal visible in the \mww model originates clearly only from R2, while the corresponding passbands are close or below the mean
level for R1. However the three redmost passbands are well above the mean level for R1, while R2 shows a flat spectrum in the red.
These differences disappear in \mwr analysis, where the both runs give compatible flat-within-uncertainties spectrum.

\subsection{Outlier passband at 807~nm}
\label{sec:discussion.problem_pb}

A single passband centred at 807~nm  in the \dsnb dataset transmission spectrum deviates from the mean.
This signal is visible on both nights, and a detailed analysis shows that the features is smooth, but somewhat
different between the nights. We could not trace the source of the feature, but can only speculate.

First, the feature is very unlikely of astrophysical origin. Certain astrophysical phenomena, such as non-transited
star spots and flares, can affect the radius ratio estimates, but the feature would require a bright narrow-band
source. 

Second, the feature is not caused by the comparison star. We repeated the DW \dsnb analyses using absolute 
photometry without dividing with the comparison star (practical since the divide-by-white approach 
removes the common-mode systematics), but this did not affect the feature.

\subsection{Mock dataset analysis}
\label{sec:discussion.mock}

We also carry out a sensitivity (reality) checks by using two mock datasets based on the \dsnb dataset. We calculate synthethic 
light curves using the observed time stamps and the system parameter posterior medians from the direct modelling. The radius
ratios follow a saw-tooth-like pattern for the mock dataset 1, and a transmission spectrum calculated by \textsc{Exo-Transmit} for the mock
dataset 2, and the limb darkening coefficients are based on the theoretical values calculated using \ldtk. The baselines
are modelled as sums of a linear time trend and a linear airmass trend with normally distributed random coefficients, and we add
normally distributed white noise with standard deviation corresponding to the true light curve standard deviation estimates.

The radius ratio posterior densities from the mock dataset analyses agree with the true radius ratios for both DW approaches.

\section{Conclusions}
\label{sec:conclusions}

We have carried out a transmission spectroscopy analysis for WASP-80b using two OSIRIS-observed
spectroscopic time series, and a joint analysis of 27 previously observed broadband light curves 
to provide reliable parameter priors for the transmission spectroscopy analysis. 

The OSIRIS data feature strong common-mode systematics, but these can be removed using either
a divide-by-white (DW) approach or initial white-light GP modelling. We chose the divide-by-white approach,
(or, more accurately, divide-by-dataset-average) which increases the per-passband white noise levels, 
but removes any systematics-model dependencies from the common-mode systematics removal.

The transmission spectroscopy analyses were repeated modelling both datasets jointly and separately,
with or without LDTk-constructed  stellar limb darkening priors, and using either a parametric or nonparametric
(GP-based) systematics model. We chose the most conservative approach as the final analysis detailed
in this paper: a DW approach accompanied with a flexible Gaussian process systematics model
where the GP hyperparameters are marginalized over in the posterior sampling process. This was motivated
by significant nightly variations in the transmission spectrum obtained using the parametric systematics model.

Despite the nightly variations in the \mww analysis, the joint analyses are consistent with each other: the 
transmission spectrum is flat within uncertainties. Especially, we do not detect significant Na or K absorption, 
and, our results do not agree with the detection of potassium by \citet{Sedaghati2017}.

The absence of significant features does not justify detailed atmospheric modelling. However, basic
\textsc{Exo-Transmit} modelling favours a truly flat spectrum over atmosphere models with stellar or 
sub-stellar metallicity and K and Na.

Ground-based observations are prone to complex systematics, and transmission spectroscopy is carried
out at the limits of what the instruments are capable of, or were designed for. The most robust approach
for ground-based transmission spectroscopy should try to account for this by repeating the observations
several times, preferably with different instruments covering the same wavelength range. Repeated
observations do not only improve the final precision that can be reached, but also our capability to decouple
the systematics from the minute transit depth variations.

\begin{acknowledgements}
We thank the anonymous referee  for their constructive and  useful comments.
HP has received support from the Leverhulme Research Project grant RPG-2012-661. FM acknowledges the 
support of the French Agence Nationale de la Recherche (ANR), under the program ANR-12-BS05-0012 
Exo-atmos. The work has been supported by the Spanish MINECO grants ESP2013-48391-C4-2-R and 
ESP2014-57495-C2-1-R.
GC acknowledges the support by the National Natural Science Foundation of China (Grant No. 11503088) and the Natural Science Foundation of Jiangsu Province (Grant No. BK20151051).
 Based on observations made with the Gran Telescopio Canarias (GTC), installed 
in the Spanish Observatorio del Roque de los Muchachos of the Instituto de Astrofísica de Canarias, 
in the island of La Palma. The authors wish to acknowledge the contribution of Teide 
High-Performance Computing facilities to the results of this research.
TeideHPC facilities are provided by the Instituto Tecnológico 
y de Energías Renovables (ITER, SA). The authors wish to acknowledge the contribution of Glamdring
computing cluster in the Subdepartment of Astrophysics, Department of Physics, University of Oxford,
to the results of this research.
\end{acknowledgements}

\bibliographystyle{aa}
\bibliography{Mendeley_WASP-80b}

\appendix

\section{DW model fit and residuals}

\begin{figure*}
	\centering
	\includegraphics[width=\textwidth]{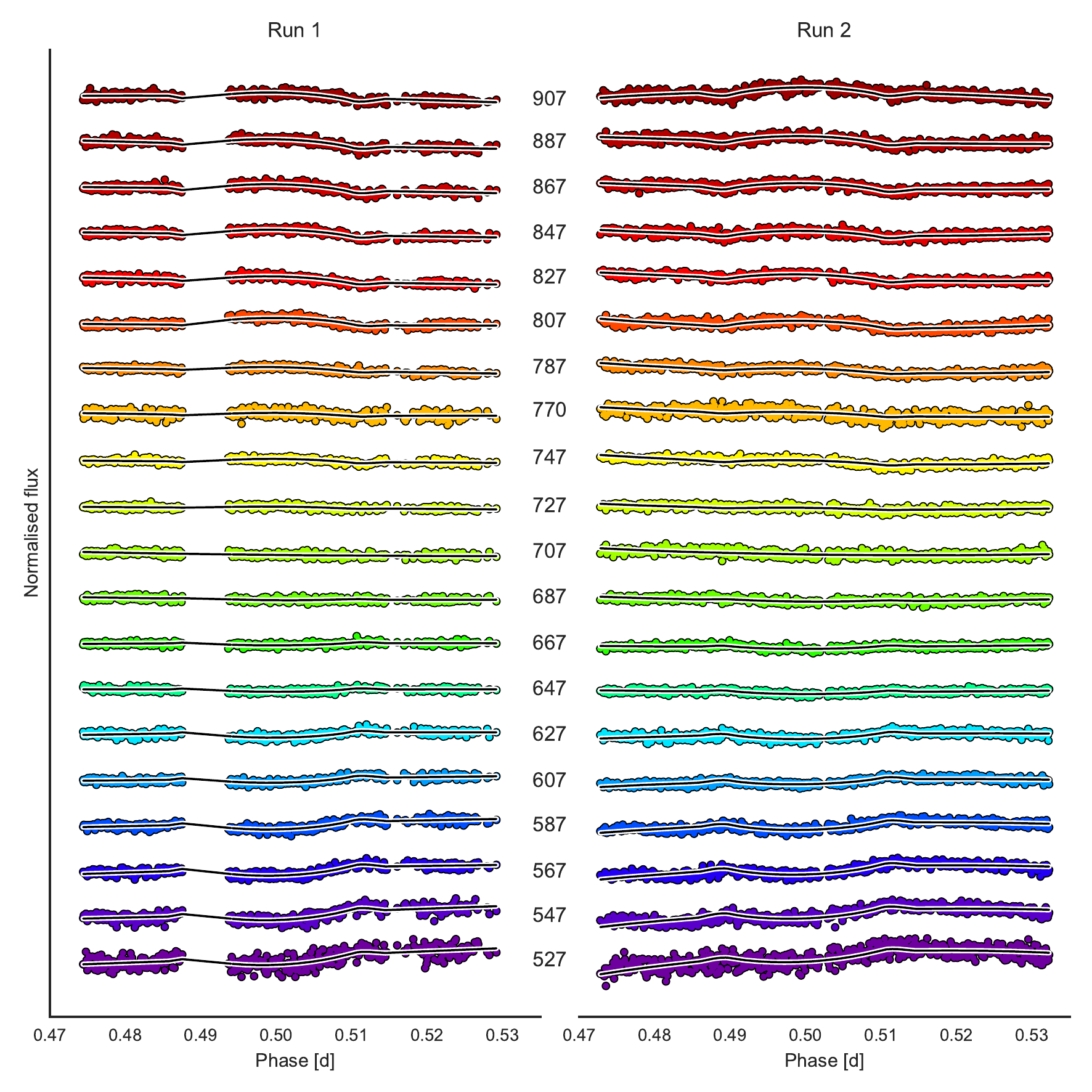}
	\caption{The \dsnb  light curves for both observing runs and the \mwr approach posterior median model  
		(black line) as a function of the phase. Passband centres are marked in nanometres.}
	\label{fig:lcs_dwr_nb}
\end{figure*}

\begin{figure*}
	\centering
	\includegraphics[width=\textwidth]{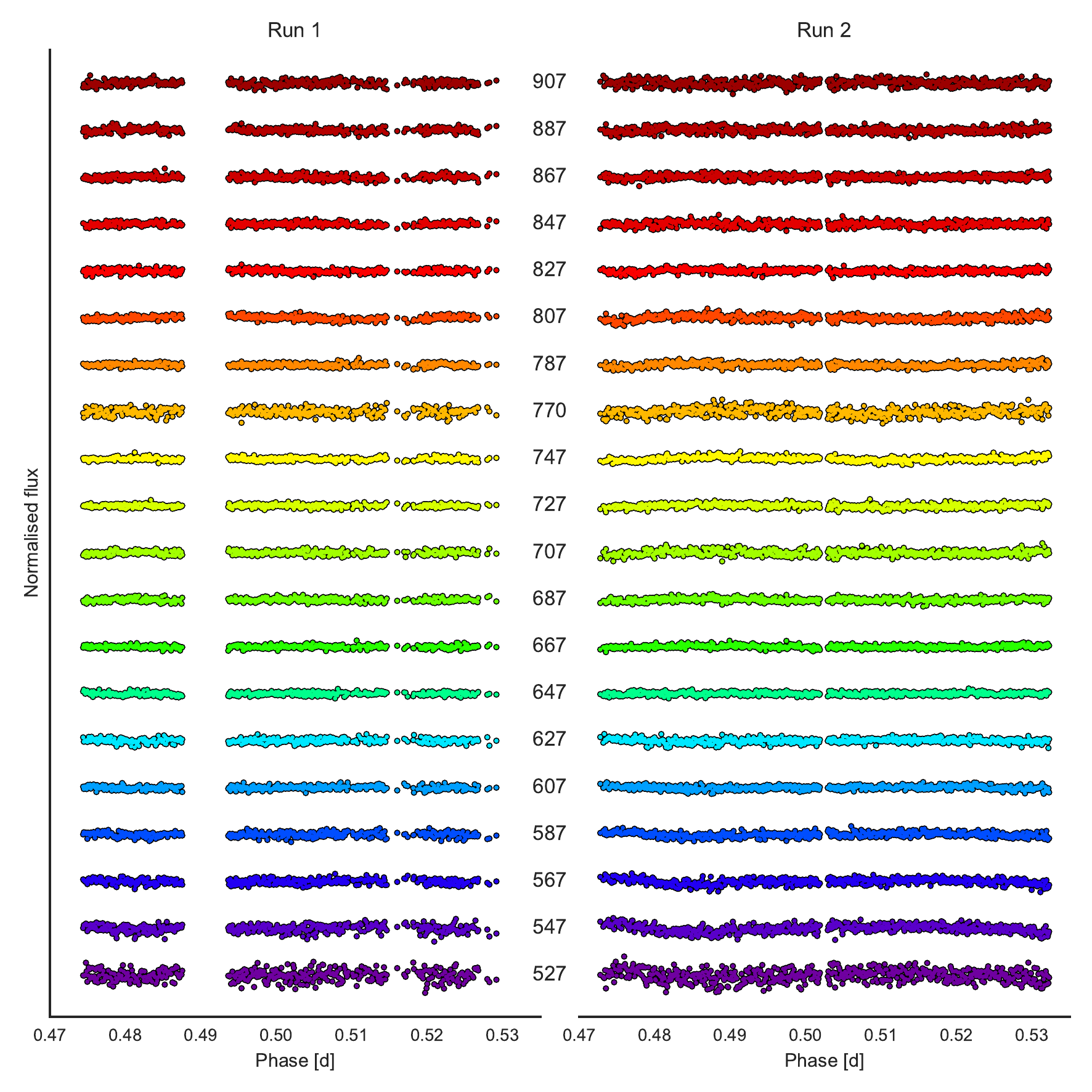}
	\caption{The \dsnb  residuals from the \mwr analysis. Passband centres are marked in nanometres.}
	\label{fig:lcs_dwr_nb_res}
\end{figure*}

\changed{Figures~\ref{fig:lcs_dwr_nb} and \ref{fig:lcs_dwr_nb_res} show the \mwr analysis model fit and residuals,
respectively.}

\end{document}